\documentclass[sigconf]{acmart}
\usepackage { amsmath , amssymb , amsthm }
\usepackage [ utf8 ]{ inputenc }
\usepackage{multirow}
\usepackage[linesnumbered,ruled,vlined]{algorithm2e}

\SetEndCharOfAlgoLine{}

\newcommand{\smalltrain}{{\sc TrainInGPU}\xspace}
\newcommand{\coa}{{\sc MultiEdgeCollapse}\xspace}
\newcommand{\lge}{{\sc LargeGraphGPU}\xspace}
\newcommand{\malgo}{{\sc \emph{Gosh}}\xspace}
\newcommand{\versex}{{\sc \emph{Verse}}\xspace}
\newcommand{\linex}{{\sc \emph{Line}}\xspace}
\newcommand{\graphvite}{{\sc \emph{Graphvite}}\xspace}
\newcommand{\mile}{{\sc \emph{Mile}}\xspace}
\newcommand{\harp}{{\sc \emph{Harp}}\xspace}

\setcopyright{acmcopyright}
\copyrightyear{2020}
\acmYear{2020}
\setcopyright{acmlicensed}\acmConference[ICPP '20]{49th International Conference on Parallel Processing - ICPP}{August 17--20, 2020}{Edmonton, AB, Canada}
\acmBooktitle{49th International Conference on Parallel Processing - ICPP (ICPP '20), August 17--20, 2020, Edmonton, AB, Canada}
\acmPrice{15.00}
\acmDOI{10.1145/3404397.3404456}
\acmISBN{978-1-4503-8816-0/20/08}
\begin{document}

\title{GOSH: Embedding Big Graphs on Small Hardware}

\author{Taha Atahan Akyildiz}
\authornote{Both authors contributed equally to this research.}
\email{aakyildiz@sabanciuniv.edu}
\affiliation{\institution{Sabanci University}}

\author{Amro Alabsi Aljundi}
\authornotemark[1]
\email{amroa@sabanciuniv.edu}
\affiliation{\institution{Sabanci University}}

\author{Kamer Kaya}
\affiliation{\institution{Sabanci University}}
\email{kaya@sabanciuniv.edu}



\begin{abstract}
In graph embedding, the connectivity information of a graph is used to represent each vertex as a point in a $d$-dimensional space. Unlike the original, irregular structural information, such a representation can be used for a multitude of machine learning tasks. Although the process is extremely useful in practice, it is indeed expensive and unfortunately, the graphs are becoming larger and harder to embed. Attempts at scaling up the process to larger graphs have been successful but often at a steep price in hardware requirements. We present \malgo, an approach for embedding graphs of arbitrary sizes on a single GPU with minimum constraints. \malgo utilizes a novel graph coarsening approach to compress the graph and minimize the work required for embedding, delivering high-quality embeddings at a fraction of the time compared to the state-of-the-art. In addition to this, it incorporates a decomposition schema that enables any arbitrarily large graph to be embedded using a single GPU with minimum constraints on the memory size. With these techniques, \malgo is able to embed a graph with over 65 million vertices and 1.8 billion edges in less than an hour on a single GPU and obtains a $93\%$ AUCROC for link-prediction which can be increased to $95\%$ by running the tool for 80 minutes.\looseness=-1
\end{abstract}

\ccsdesc[500]{Parallel computing methodologies~Parallel algorithms}
\ccsdesc[300]{Parallel computing methodologies~Shared memory algorithms}
\ccsdesc[500]{Mathematics of computing~Discrete mathematics}
\ccsdesc[300]{Mathematics of computing~Graph theory}
\ccsdesc[100]{Mathematics of computing~Graph algorithms}

\keywords{Graph embedding, GPU, parallel graph algorithms, graph coarsening, link prediction}
\maketitle

\section{Introduction}

Graphs are widely adopted to model the interactions within real-life data such as social networks, citation networks, web data, etc. Recently, using machine learning~(ML) tasks such as link prediction, node classification, and anomaly detection on graphs became a popular area with various applications from different domains. 
The raw connectivity information of a graph, represented as its adjacency matrix, does not easily lend itself to be used in such ML tasks; regular $d$-dimensional representations are more appropriate for learning valid correlations between graph elements. Unfortunately, the connectivity information does not have such a structure. Recently, there has been a growing interest in the literature in the \emph{graph embedding} problem which focuses on representing the vertices of a graph as $d$-dimensional vectors while embedding its structure into a $d$-dimensional space.

Various graph embedding techniques~\cite{verse18, LINE, deepwalk, node2vec} have been proposed in the literature. 
However, these approaches do not usually scale to large, real-world graphs. For example, \versex~\cite{verse18}, \emph{deepwalk}~\cite{deepwalk}, \emph{node2vec}~\cite{node2vec} and \linex~\cite{LINE} require hours of CPU training, even for small- and medium-scale graphs. 
Although these approaches can be parallelized, a multi-core CPU implementation of \versex takes more than two hours on 16 CPU cores for a graph with 2 million vertices and 20 million edges. 
There are other attempts to increase graph embedding performance. 
\harp~\cite{HARP} and \mile~\cite{mile18} apply graph coarsening, a process in which a graph is compressed into smaller graphs, to make the process faster, but they do not have a parallel implementation. 
Accelerators such as GPUs can be used to deal with large-scale graphs. However, to the best of our knowledge, the only GPU-based tool in the literature is \graphvite. 
Although \graphvite is faster than the CPU counterparts, its use is limited by the device memory and it cannot embed large graphs with a single GPU.\looseness=-1 

In this paper, we present \malgo\footnote{The code is publicly available at \url{https://github.com/SabanciParallelComputing/GOSH}}; an algorithm that performs parallel coarsening and many-core parallelism on GPU for fast embedding. The tool is designed to be fast and accurate and to handle large-scale graphs even on a single GPU. However, it can easily be extended to the multi-GPU setting. \malgo performs the embedding in a multilevel setting by using a novel, parallel coarsening algorithm which can shrink graphs while preserving their structural information and trying to avoid giant vertex sets during coarsening. With coarsening, the initial graph is iteratively shrunk into multiple levels. Then, starting from the smallest graph, unsupervised training is performed on the GPU. The embedding obtained from the current level is directly copied to the next one by using the coarsening information obtained from the above level. The process continues with the expanded embedding for the next level until the original graph is processed on the GPU and the final embedding is obtained. The contributions can be summarized as follows:
\begin{itemize}
    \item To the best of our knowledge, the only GPU-based graph embedding tool, \graphvite, cannot handle embedding large graphs on a single GPU when the total size of the embedding is larger than the total available memory of the GPU. On the contrary, \malgo applies a smart decomposition, update scheduling, and synchronization to perform the embedding even when the memory requirement exceeds the memory available on a single GPU.  
   
    \item Thanks to multilevel coarsening and smart work distribution across levels, \malgo's embedding is ultra-fast. 
    For instance, on the graph {\tt com-lj}, \graphvite, a state-of-the-art GPU-based embedding tool, spends around $11$ minutes to reach $98.33\%$ AUCROC score. On the same graph, \malgo spends only $2.5$ minutes and obtains $98.46\%$ AUCROC score. Furthermore, based on the numbers given in~\cite{graphvite19}, the embedding takes more than 20 hours on 4 Tesla P100 GPUs for a graph of 65 million vertices and 1.8 billion edges (the link prediction scores are not reported in~\cite{graphvite19} for this graph). On a single Titan X GPU, \malgo reaches $95\%$ AUCROC score within 1.5 hours.\looseness=-1 

    \item The dimension, i.e., the number of features, used for the embedding process can vary with respect to the application. For different $d$ values, the best possible GPU-implementation, which utilizes the device cores better, also differs. For different $d$ values, \malgo performs different parallelization strategies to further increase the performance of embedding especially for small $d$ values.\looseness=-1 
   
    \item The multilevel setting for graph embedding has been previously applied by \mile~\cite{mile18} and \harp~\cite{HARP}. Since CPU-based embedding takes hours, the coarsening literature suggests that the time required for coarsening is negligible in comparison. In this work, we show that a parallel coarsening algorithm is necessary since \malgo is orders of magnitude faster than CPU-based approaches. To overcome this bottleneck, we propose an efficient and parallel coarsening algorithm which is empirically much faster than that of \mile and \harp. 
\end{itemize}

The rest of the paper is organized as follows: In Section~\ref{sec:not}, the notation used in the paper is given. 
Section~\ref{sec:gosh} describes \malgo in detail including the coarsening algorithm and the techniques designed and implemented to handle large graphs. 
The experimental results are presented in Section~\ref{sec:exp} and the related work is summarized comparatively in Section~\ref{sec:rel}. 
Section~\ref{sec:con} concludes the paper.\looseness=-1 

\section{Notation and Background}\label{sec:not}
A graph $G = (V, E)$ has $V$ as the set of nodes/vertices and $E \subseteq (V \times V)$ as the set of edges among them. For undirected graphs, an edge is an unordered pair while in directed graphs, the order is significant. 
An embedding of a graph $G = (V,E)$ is a $|V| \times d$ matrix ${\mathbf M}$, where $d$ is the dimension of the embedding. The vector ${\mathbf M}[i]$ corresponds to a vertex $i \in V$ and each value $j$ in the vector ${\mathbf M}[i][j]$ captures a different feature of vertex $i$. The embedding of a graph can be used in many machine learning tasks such as link prediction \cite{linkprediction},  node classification \cite{deepwalk} and anomaly detection \cite{anomaly_detection}.

There are many algorithms in the literature for embedding the nodes of a graph into a d-dimensional space.
\malgo implements the embedding method of \versex~\cite{verse18}; a method which, in addition to having fast run-time and low memory overhead, is highly generalizable as it can produce embeddings that reflect \emph{any} vertex-to-vertex similarity measure $Q$.
To elaborate, this approach defines two distributions for each vertex $v$: The first, $sim_{Q}^v$, is obtained from the similarity values between $v$ and every other vertex in $G$ computed based on $Q$. The second, $sim_{E}^v$, is derived from the embedding by using the cosine similarities of $v$'s embedding vector and those of every other vertex in $G$. A soft-max normalization is applied to these values as a post-processing step so they sum up to $1$. The problem then becomes the minimization of the Kullback-Leibler (KL) divergence between the two distributions for every vertex. In this paper, we choose $Q$ to be the adjacency similarity measure described in~\cite{verse18}\looseness=-1 . 


\newcommand{\updateembedding}{{\sc UpdateEmbedding}}

\begin{algorithm}
\KwData{$\mathbf{M}[v]$, $\mathbf{M}[sample]$, $b$, 
        $lr$}
 \KwResult{$\mathbf{M}[v]$, $\mathbf{M}[sample]$}
 $score \leftarrow b - \sigma(\mathbf{M}[v] \odot \mathbf{M}[sample]) \times lr$ \;
 $\mathbf{M}[v] \leftarrow \mathbf{M}[v] + \mathbf{M}[sample] \cdot score$\;
 $\mathbf{M}[sample] \leftarrow \mathbf{M}[sample] + \mathbf{M}[v] \cdot score$\;

\caption{\updateembedding}\label{alg:update}
\end{algorithm}

The training procedure employs {\em Noise Contrastive Estimation} for convergence of the objective above as described in~\cite{verse18}. The process trains a logistic regression classifier to separate vertex samples drawn from the empirical distribution $Q$ and vertex samples drawn from a noise distribution $N$, with the corresponding embedding vectors being the parameters of this classifier. More precisely, all the vertices are processed a total number of $e$ times, i.e., {\em epochs}, where processing a vertex $v \in V$ consists of drawing a single positive sample $u$ from $sim_{Q}^v$ and $n_s$ negative samples $s_{1}, s_{2}, \dots, s_{n_s}$ from $N$. In the meantime, logistic regression is used to minimize the negative log-likelihood of observing $u$ and not observing $s_{1}, s_{2}, \dots, s_{n_s}$ by updating the corresponding embedding vectors of $v$ and all the other samples. A single update is shown in Algorithm~\ref{alg:update}, where $\mathbf{M}[v]$ is the embedding vector of $v$, $\mathbf{M}[sample]$ is the embedding vector of the sample, $b$ is a binary number that is $1$ if the sample is positive (drawn from $Q$) or negative (drawn from $N$), $\odot$ is the dot-product operation, $\sigma$ is the sigmoid function and $lr$ is the learning rate of the classifier.

The notation used in the paper is given in Table~\ref{tab:notation}. 

\begin{table}
  \caption{Notation used in the paper.}
  \label{tab:notation}
  \scalebox{0.88}{
  \begin{tabular}{ll}
    \toprule
    \textbf{Symbol} & \textbf{Definition}\\
    \midrule
    $G_0 = (V_0, E_0)$ & The original graph to be embedded.\\
    $G_{i} = (V_{i}, E_{i})$ & Represents a graph, which is coarsened $i$ times.\\
    $\Gamma^+(u)$ & The set of outgoing neighbors of vertex $u$. \\
    $\Gamma^-(u)$ & The set of incoming neighbors of vertex $u$. \\
    $\Gamma(u)$ & Neighborhood of $u$, i.e., $\Gamma_{G_{i}}^+(u) \bigcup \Gamma_{G_{i}}^-(u)$.\\\hline
    $d$ & $\#$ features per vertex, i.e., dimension of the embedding.\\
    $n_s$ & $\#$ negative samples per vertex.\\
    $\sigma$ & Sigmoid function.\\
    $sim_{m}$ & Similarity metric used in training.\\ 
    $e$ & Total number of epochs that will be performed\\
    $lr$ & Learning rate.\\\hline
    $D$ & Total amount of coarsening levels. \\
    ${\mathcal G}$ & The set of coarsened graphs created from a graph $G = G_0$. \\
    $p$ & Smoothing ratio for epoch distribution.\\
    $e_i$ & $\#$ epochs for coarsening level $i$.\\
    ${\mathbf M}_{i}$ & Embedding matrix obtained for $G_{i}$.\\
    ${\mathcal M}$ & The set of mappings used in coarsening.\\
    $map_i$ & Mapping information from $G_{i-1}$ to $G_{i}$. \\\hline
    ${\mathcal V}_{i}$ & The partitioning of vertex set $V_i$. \\
    ${\mathcal P}_{i}$ & The partitioning of embedding matrix ${\mathbf M}_i$. \\
    $K_i$ & $\#$ parts in ${\mathcal V}_{i}$. \\
    ${P}_{GPU}$ & $\#$ embedding parts to be placed on the GPU. \\
    $S_{GPU}$ & $\#$ sample pools to be placed on the GPU. \\
    $B$ & $\#$ positive samples per vertex in a single sample pool. \\
  \bottomrule
\end{tabular}}
\end{table}

\begin{figure}
    \centering
    \includegraphics[width=0.9\linewidth]{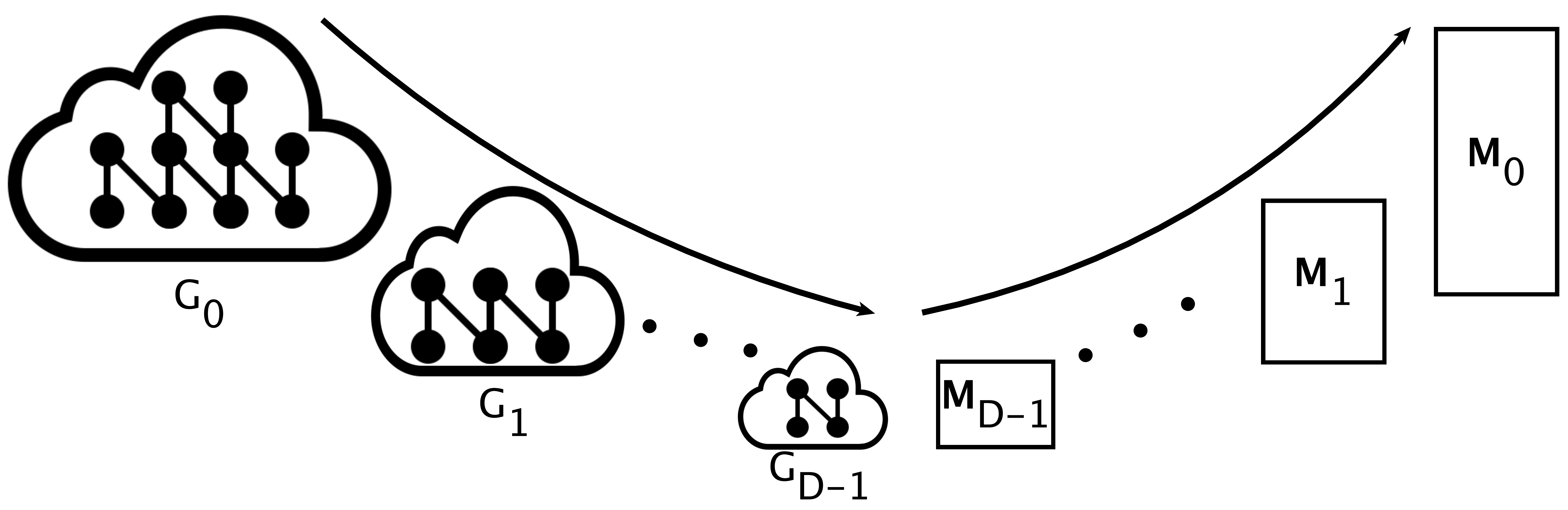}
    \caption{{Multilevel embedding performed by \malgo: first, the coarsened set of graphs are generated. Then, the embedding matrices are trained until ${\mathbf M}_0$ is obtained.}}
    \label{fig:vcycle}
    \Description[A graph is coarsened from the original to smaller and smaller graphs then embedded and expanded]{The original graph is coarsened down into a smaller graph, and then that graph is coarsened, and so on. The smallest graph is then embedded, and its embeddings are projected onto the higher graph, and then that graph is embedded and that continues upwards until the original graph is embedded.}
\end{figure}

\section{Embedding on Small Hardware}\label{sec:gosh}

This section is organized as follows: First, a high-level explanation of GOSH is provided. Then, Section 3.1 describes the GPU implementation for the embedding process in detail. Following the embedding process, in Section 3.2, a new coarsening approach is introduced and the parallel implementation details are provided. Finally, Section 3.3 describes the partitioning schema which enables \malgo to handle graphs that do not fit on the GPU memory.

Given a graph $G_{0}$, \malgo, shown in Algorithm~\ref{alg:malgo}, computes the embedding matrix ${\mathbf M}_{0}$. This is done in two stages; 
\begin{enumerate}
    \item creating a set ${\mathcal G} = \{G_0, G_1, \ldots, G_{D-1}\}$ of graphs coarsened in an iterative manner (as in the left of Figure~\ref{fig:vcycle}) where one or more than one nodes in $G_{i-1}$ are represented by a super node in $G_{i}$ (Line~\ref{ln:malgo:coarsening}),
    \item starting from $G_{D - 1}$, training the embedding matrix ${\mathbf M}_{i}$ for the  graph $G_i$ and projecting it to $G_{i-1}$ to later train ${\mathbf M}_{i-1}$~(as in the right of Figure~\ref{fig:vcycle})~(Lines~\ref{ln:malgo:bfor}-~\ref{ln:malgo:efor}).
\end{enumerate}
\noindent The training process is repeated until ${\mathbf M}_0$ is obtained. To obtain ${\mathbf M}_{i-1}$ from ${\mathbf M}_{i}$ the mapping information of ${\mathbf G}_{i-1}$ is used, where $M_{i}[u] = M_{i-1}[v]$ iff $u \in V_{i}$ is a super node of $v \in V_{i-1}$.

\malgo provides support for large-scale graphs for which the memory footprint of the training exceeds the device memory. Even for practical sizes, e.g., $|V| = 128$M and $d = 128$, the number of entries in the matrix is approximately $16$G. With double precision, one needs to have $128$GB memory on the device to store the entire matrix. 
For each $G_{i}$, \malgo initially checks if both $G_{i}$ and ${\mathbf M}_{i}$ can fit in the GPU (Line~\ref{ln:malgo:ifstatment}). If so, it proceeds by copying $G_{i}$ and the projection of ${\mathbf M}_{i}$ to the GPU and carrying out the embedding of $G_{i}$ in a single step (Lines ~\ref{ln:malgo:copy}-\ref{ln:malgo:smalltrain}). Otherwise, it generates positive samples in CPU and carries out the embedding of $G_{i}$ by copying the relative portions of the samples and ${\mathbf M}_{i}$ and training the graph in batches~(Line ~\ref{ln:malgo:bg}).

Using multilevel coarsening arises an interesting problem; let $e$ be the total number of epochs one performs on all levels. With a naive approach, one can distribute the epochs evenly to each level. However, when more epochs are reserved for $G_i$s in the lower levels, the process will be faster. To add, the corresponding embedding matrices will have a significant impact on the overall process as they are projected to further levels. On the other hand, when more epochs are reserved for the higher levels, i.e., larger $G_i$s, the embedding is expected to be more fine-tuned. So the question is {\em how to distribute the epoch budget to the levels}. Based on our preliminary experiments, \malgo employs a mixed strategy; a portion, $p$, of the epochs are distributed uniformly and the remaining $(e \times (1-p))$ epochs are distributed geometrically. That is, training at level $i$ uses $e_i = e/D + e'_i$ epochs where $e'_i$ is half of $e'_{i+1}$. 
 The value $p$ is called the {\em smoothing ratio} and is left as a configurable parameter for the user to establish an interplay between the performance and accuracy.\looseness=-1

Another parameter that has a significant impact on embedding quality is the learning rate. 
For multilevel embedding, a question that arises is {\em how to set the learning strategy for each level}. 
In \malgo, we use the same initial learning rate for each level, i.e., for the training of each ${\mathbf M}_i$ and decrease it after each epoch. 
That is, the learning rate for epoch $j$ at the $i$th level is equal to $lr \times \max\left(1 - \frac{j}{e_i}, 10^{-4}\right)$. 

%

%
\begin{algorithm}
\KwData{$G_{0}$, $n_s$, $lr$, 
        $lr_d$, 
        $p$, $e$, $threshold$, $P_{GPU}$, $S_{GPU}$, $B$}
 \KwResult{${\mathbf M}$}
 ${\mathcal G} \leftarrow $\coa$(G_{0}, threshold)$\;\label{ln:malgo:coarsening}
  Randomly initialize ${\mathbf M}_{D-1}$\;
 \For{$i$ from $D-1$ to $1$}
 {\label{ln:malgo:bfor}
    $e_i \leftarrow$ {\sc calculateEpochs}($e$, $p$, $i$)\; 
    \If{$G_{i}$ and ${\mathbf M}_{i}$ fits into GPU}
    {\label{ln:malgo:ifstatment}
        {\sc CopyToDevice}($G_{i}$, ${\mathbf M}_{i}$)\;\label{ln:malgo:copy}
        ${\mathbf M}_{i} \leftarrow$ \smalltrain($G_{i}$, ${\mathbf M}_i$, $n_s$, $lr$, $lr_d$, $e_i$)\;\label{ln:malgo:smalltrain}
    }
    \Else
    {
         ${\mathbf M}_{i} \leftarrow$ \lge($G_{i}$, ${\mathbf M}_i$, $n_s$, $lr$, $lr_d$, $e_i$, \\\hspace*{25ex}$P_{GPU}$, $S_{GPU}$, $B$)\;\label{ln:malgo:bg}
    }
        ${\mathbf M}_{i-1} \leftarrow$ {\sc ExpandEmbedding}(${\mathbf M}_{i}$, $map_{i-1}$)\;\label{ln:malgo:expand}

}\label{ln:malgo:efor}
{\bf return $\mathbf{M}_0$}\;
\caption{\malgo}\label{alg:malgo}
\end{algorithm}

\subsection{Graph embedding in \malgo}

\malgo implements a GPU parallel, lock-free learning step for the embedding algorithm. As mentioned above, following {\em VERSE}, we use an SGD-based optimization process for training. 
As shown in~\cite{hogwild}, a lock-free implementation of SGD, which does not take the race conditions, i.e., simultaneous updates, into account, does not have a significant impact on the learning quality of the task on multi-core processors. However, our preliminary experiments show that on a GPU, where millions of threads are being executed in parallel, such race conditions significantly deteriorate the quality of the embedding. For \malgo, we follow a slightly more restricted implementation which is still not race-free. 

To reduce the impact of race conditions, we synchronize the epochs and ensure that no two epochs are processed at the same time. For an epoch to train ${\mathbf M}_i$, \malgo traverses $V_{i}$ in parallel by assigning a single vertex to a single GPU-warp. For a single (source) vertex, multiple positive and negative samples are processed one after another and updates are performed as in Algorithm~\ref{alg:update}. With this implementation and vertex-to-warp assignment, a vertex $v \in V_{i}$ cannot be a source vertex for two concurrent updates but it may be (positively or negatively) sampled by another vertex while the warp processing $v$ is still active. Similarly, $v$ can be sampled by two different source vertices at the same time. Hence, the reads/writes on ${\mathbf M}_i[v]$ are not completely race free. However, according to our experiments, synchronizing the epochs and samples for the same source vertex is enough to robustly perform the embedding process. 

As shown in Algorithm~\ref{alg:EA}, the positive and negative samples are generated in the GPU. For a source vertex $v \in V_i$, the positive sample $u \in V_i$ is chosen from $\Gamma_{G_{i}}(v)$. 
Negative samples, on the other hand, are drawn from a noise distribution as mentioned in Section~\ref{sec:not} which we model as a uniform random distribution over $V_{i}$. 
After each sampling, the corresponding update is performed by using the procedure in Algorithm~\ref{alg:update}. 

\begin{algorithm}[htbp]
 \KwData{$G_{i}$, ${\mathbf M}_i$, $n_s$, $lr$, $e_i$}
 \KwResult{${\mathbf M}_{i}$}
  \For{$j = 0$ to $e_i - 1$}{ 
  $lr' \leftarrow lr \times max \left(1 - \frac{j}{e_i}, 10^{-4}\right)$\;
  \tcc{Each $src$ below is assigned to a GPU warp}
  \For{$\forall src \in V_i$ \bf in parallel} {
    $u \leftarrow$ {\sc GetPositiveSample}($G_i$)\;
     \updateembedding($\mathbf{M}_i[src]$, ${\mathbf M}_i[u]$, $1$,  $lr'$ )\; 
      \For{$k = 1$ to $n_s$}{
         $u \leftarrow$ {\sc GetNegativeSample}($G_i$)\;
         \updateembedding($\mathbf{M}_i[src]$, ${\mathbf M}_i[u]$, $0$, $lr'$)\;
       }
    }
 }
\caption{\smalltrain}\label{alg:EA}
\end{algorithm}

During the updates for a source vertex $src$, the threads in the corresponding warp perform $(1 + n_s) \times d$ accesses to ${\mathbf M}_{i}[src]$. For large values of $n_s$ and $d$, performing this many global memory accesses dramatically decreases the performance. To mitigate this, before processing $src$, \malgo copies ${\mathbf M}_{i}[src]$ from global to shared memory. Then for all positive and negative samples, the reads and writes for the source are performed on the shared memory. Finally, ${\mathbf M}_{i}[src]$ is copied back to global memory. On the other hand, for the sampled $u$ vertices, ${\mathbf M}_{i}[u]$ is always kept in global memory since each entry is read and written only once. To perform coalesced accesses on ${\mathbf M}_{i}[u]$, the reads and writes are performed in a round-robin fashion. That is ${\mathbf M}_{i}[u][j + (32 \times k)]$ is accessed by thread $j$ at the $k$th access where $32$ is the number of threads within a warp.  

\subsubsection{Embedding for small dimensions} \label{sec:small_dimensions}

Assuming a warp contains 32 threads, when $d \leq 16$ dimensions are used for embedding, a single source vertex does not keep all the threads in a warp busy. In this case, $32 - d$ warp threads remain idle which yields to the under-utilization of the device. To tackle this problem, we integrate a specialized implementation for small dimension embedding. We set the number of threads responsible for a source vertex as the smallest multiple of $8$ larger than or equal to $d$, i.e., $8$ or $16$. Hence, depending on $d$, we can assign $2$ or $4$ vertices to a single warp. 

\subsection{Graph coarsening}\label{s_coarsening}
%

 
\malgo employs a fast algorithm to keep the structural information within the coarsed graphs while maximizing the coarsening {\em efficiency} and {\em effectiveness}. Coarsening efficiency at the $i$th level is measured by the rate of shrinking defined as ${(|V_{i - 1}| - |V_{i}|)}/{|V_{i - 1}|}$. On the other hand, the effectiveness is measured in terms of its embedding quality compared to other possible coarsenings of the same graph embedded with the same parameters. 
We adapt an agglomerative coarsening approach, \coa, which generates vertex clusters in a way similar to the one used in~\cite{HARP}. Given $G_i = (V_i, E_i)$, the vertices in $V_i$ are processed one by one. If $v$ is not marked, it is marked, and mapped to a cluster, i.e., a new vertex in $V_{i+1}$ and its edges are processed. If an edge $(v,u) \in E_{i}$, where $u$ is not marked, $u$ is added to $v$'s cluster. Then, all of the vertices in $v$'s cluster are shrunk into a {\em super} vertex $v_{sup} \in G_{i + 1}$.\looseness=-1

\coa preserves both the first- and second-order proximites~\cite{LINE} in a graph. The former measures the pairwise connection between vertices, and the latter represents the similarity between vertices' neighborhoods. It achieves that by collapsing vertices that belong to the same neighborhood around a local, hub vertex.
However, if this process is handled carelessly, two, giant hub vertices can be merged. We observed that this degrades the effectiveness and efficiency of the coarsening. The effectiveness degrades since the structural equivalence is not preserved in the lower levels of the coarsening, where most of the vertices are represented by a small set of super vertices. Furthermore having a small set of giant supers inhibits the graph from being coarsened further, resulting in an insufficient efficiency. To mitigate this, a new condition for matching is introduced to the algorithm, where $u \in V_{i}$ can not be put into the cluster of $v \in V_i$ if $|\Gamma_{G_{i}}(u)|$ and $|\Gamma_{G_{i}}(v)|$ are both larger than $\frac{|E_{i}|}{|V_{i}|}$. Consequently, assuming that the hub vertices will have a higher degree than the density of $G_{i}$, two of them can no longer be in the same cluster. Our preliminary experiments show that this simple rule has a significant effect on both the efficiency and the effectiveness of the coarsening.

As mentioned above, when a vertex is marked and added to a cluster, its edges are not processed further and it does not contribute to the coarsening. Performing the coarsening with an arbitrary ordering may degrade the efficiency since large vertices can be locked by the vertices with small neighborhoods. Hence, when an edge $(u, v) \in E_i$ is used for coarsening for a hub-vertex $v \in V_i$, to maximize efficiency, we prefer $u \in V_i$ to be inserted in to the cluster of origin $v$. To provide this, an ordering is procured by sorting the vertices with respect to their neighborhood size and this ordering is used during coarsening. This ensures processing vertices with a higher degree before the vertices with smaller neighborhoods and this results in a substantial increase in the coarsening efficiency.

The details of the coarsening phase are given in Algorithm~\ref{alg:MEC}. The algorithm takes an uncoarsened graph $G = G_{0}$ and returns the set of coarsened graphs ${\mathcal G}$ along with the mapping information to be used to project the embedding matrices ${\mathcal M}$. ${\mathcal G}$ and ${\mathcal M}$ are initialized as $\{G_{0}\}$ and empty set, respectively. Starting from $i = 0$, the coarsening continues until a graph $G_{i+1}$ with less than $threshold$ vertices is generated. As mentioned above, first the vertices in $G_i$ are sorted with respect to their neighborhood sizes. Then the coarsening is performed and a smaller $G_{i+1}$ is generated. We also store the mapping information $map_i$ used to shrink $G_i$ to $G_{i+1}$. This will be used later to project the embedding matrix $M_{i+1}$ obtained for $G_{i+1}$ to initialize the matrix $M_i$ for $G_i$. 
To add, $threshold = 100$ is used for all the experiments in the paper which is the default value for \malgo. 

\begin{algorithm}[htbp]
 \KwData{$G_0 = (V_0, E_0)$, $threshold$}
 \KwResult{${\mathcal G}$, ${\mathcal M}$}
 ${\mathcal G} \leftarrow \{G_{0}\}$, ${\mathcal M} \leftarrow \emptyset$, $i \leftarrow 0$\;
 \While{$|V_{i}| > threshold$}{
    $order \leftarrow$ {\sc Sort}($G_{i}$)\;\label{ln:sort}
    \lFor{$v \in V_i$} {$map_i[v] \leftarrow - 1$}
    $\delta \leftarrow |E_i| / |V_i|$\;
    $cluster \leftarrow 0$\;
    \For{$v$ in $order$} { \label{ln:fors}
        \If{$map_i[v] = -1$} {
            $map_i[v] \leftarrow cluster$\;\label{ln:map1}
            $cluster \leftarrow cluster + 1$\;
            \ForEach{$(v, u) \in E_i$} {
                    \If{$|\Gamma_{G_i}(v)| \leq \delta$ or $|\Gamma_{G_i}(u)| \leq \delta$} {
                                    \If{$map_i[u] = -1$} {
                        $map_i[u] \leftarrow map_i[v]$\;\label{ln:map2}
                    }
                }
            } 
        }
    }\label{ln:fore}
    $G_{i+1} \leftarrow$ {\sc Coarsen}($G_{i}, map_{i}$)\;\label{ln:coarse}
    ${\mathcal G} \leftarrow {\mathcal G} \cup \{G_{i+1}\}$, ${\mathcal M} \leftarrow {\mathcal M} \cup \{map_{i}\}$, $i \leftarrow i + 1$\;
 }
 \caption{\coa}\label{alg:MEC}
\end{algorithm}

\subsubsection{\it{Complexity analysis:}} 
All the algorithms, coarsening and embedding, use the Compressed Sparse Row~(CSR) graph data structure. In CSR, an array, $adj$ holds the neighbors of every vertex in the graph consecutively. It is a list of all the neighbors of vertex $0$, followed by all the neighbors of vertex $1$, and so on. Another array, $xadj$, holds the starting indices of each vertex's neighbors in $adj$, with the last index being the number of edges in the graph. In other words, the neighbors of vertex $i$ are stored in the array $adj$ from $adj[xadj[i]]$ until $adj[xadj[i+1]]$. 

\coa has three stages; sorting~(line \ref{ln:sort}), mapping~(lines \ref{ln:fors}--\ref{ln:fore}) and coarsening~(line \ref{ln:coarse}). A \emph{counting sort} is implemented for the first stage with a time complexity of $\mathcal{O}(|V| + |E|)$. For mapping, the algorithm traverses all the edges in the graph. This has a time complexity of $\mathcal{O}(|V| + |E|)$. Finally, coarsening the graph requires sorting the vertices with respect to their mappings and going through all the vertices and their edges within the CSR, which also has a time complexity of $\mathcal O(|V| + |E|)$.\looseness=-1

\subsubsection{Parallelization:} \label{s_coarsening_parallel}
When the embedding is performed on the CPU, as the literature suggests, embedding dominates the total execution time. However, with fast embedding as in \malgo, this is not the case. Thus, we parallelize the coarsening on the CPU.

In a parallel coarsening with $\tau$ threads, one can simply traverse $V_{i}$ in parallel and perform the mapping with no synchronization. However, this creates race conditions and inconsistent coarsenings.
To avoid race conditions, we use a lock per each entry of $map_i$. To update $map_{i}[v]$ and $map_{i}[u]$ as in lines~\ref{ln:map1} and~\ref{ln:map2}, the thread first tries to lock $map_{i}[v]$ and $map_{i}[u]$, respectively. If the lock is obtained, the process continues. Otherwise, the thread skips the current candidate and continues with the next vertex. One caveat is the update on the counter $cluster$. Hence, instead of using a separate variable for super vertex ids, the parallel version uses the hub-vertex id for mapping. That is $map_i[v]$ is set to $v$ unlike line \ref{ln:map1} of the sequential algorithm. Note that with this implementation, $map_i$ does not provide a mapping to actual vertex IDs in $G_{i+1}$. This can be fixed in $\mathcal{O}(|V|)$ time via sequential traversals of the $map_i$ array, which first detect/count the vertices that has $map_i[v] = v$ and reset the $map_i$ values for all.   
%
%

The parallel coarsened graph construction is not straightforward. After the mapping, the degrees of the (super) vertices in $G_{i+1}$ are not yet known. To alleviate that, we allocate a private $E^j_{i+1}$  region in the memory to each thread $t_j$, $1 \leq j \leq \tau$. These threads create the edge lists of the new vertices on these private regions which are then merged on a different location of size $|E_{i+1}|$. To do that first a sequential scan operation is performed to find the region in $E_{i+1}$ for each thread. Then the private information is copied to $E_{i+1}$.  

An important problem that needs to be addressed for all the steps above is load imbalance. Since the degree distribution on the original graph can be skewed and becomes more skewed for the coarsened graphs, a static vertex-to-thread assignment can reduce the performance. Hence \malgo uses a dynamic scheduling strategy, which uses small batch sizes for all the steps above.
\subsection{Handling large graphs}\label{meth_lg}
One of the strongest points of \malgo is the ability to quickly generate a high-quality embedding of a large-scale graph on a single GPU. Here we present the techniques used when the graph and the embedding matrix do not fit in the GPU memory. The base algorithm for \malgo requires storing $d \times |V|$ and $(|V| + 1) + |E|$ entries for the matrix and the graph, respectively, for a graph $G =(V,E)$. For large graphs with millions of vertices, a single GPU is unable to store this data.\looseness=-1

We mitigate the bottleneck of storing ${\mathbf M}_{i}$ on the GPU by using a partitioning schema similar to \cite{graphvite19,pbg19}, in which ${\mathbf M}_{i}$ is partitioned and embedding is carried out on the sub-matrices instead of the entirety of ${\mathbf M}_{i}$. Formally, we partition $V_{i}$ into $K_{i}$ disjoint subsets of vertices ${\mathcal V}_{i} = \{V_{i}^{0}, V_{i}^{1}, \dots, V_{i}^{K_{i}-1}\}$. 
Let ${\mathcal P}_{i} = \{{\mathbf M}_{i}^{0}, {\mathbf M}_{i}^{1}, \dots, {\mathbf M}_{i}^{K_{i}-1}\}$ be the sub-matrices of ${\mathbf M}_{i}$ corresponding to the vertex sets in ${\mathcal V}_{i}$.
With partitioning, embedding $G_{i}$ becomes the process of moving the sub-matrices in ${\mathcal P}_{i}$ to the GPU, carrying out the training on these parts, and switching them out for the next sub-matrices, and so on.

In order to carry out the embedding correctly, all possible negative samples, chosen from  $V_{i} \times V_{i}$, must be able to be processed. To do this, \malgo handles the embedding in rotations. During a rotation there will always be a point in time when ${\mathbf M}_{i}^{j}$ and ${\mathbf M}_{i}^{k}$ are together in the GPU for all $0 \leq j < k < K$. When this happens, $B$ positive and $B \times n_s$ negative samples are chosen from $V_{i}^{j}$ (or $V_{i}^{k}$) for every vertex in $V_{i}^{k}$ (or $V_{i}^{j}$).
This way, in each rotation, we run a total of at most $B \times K_{i}$ updates on every vertex, which makes a full rotation (almost) equivalent to $B \times K_{i}$ epochs. We use the term {\em almost} since a vertex $v \in V_{i}^{j}$ may not have a neighbor in $V_{i}^{k}$.
In this case, no positive updates are performed for $v$ when$V_{i}^{j}$ and $V_{i}^{k}$ are on the device. Hence, running $\frac{e_i}{B \times K_{i}}$ rotations is (almost) equivalent to running $e_i$ epochs for the embedding. We set $B = 5$ as the default value in \malgo and experiment with different values to see its impact on embedding quality and performance.\looseness=-1

Although partitioning the embedding matrix solves the first memory bottleneck, storing $G_{i}$ on the GPU can still be problematic since this will leave less space for the sub-matrices. This is why we opt not to store $G_{i}$ on the GPU. The positive samples are chosen on the host and only when required, they are sent to the GPU. 
Negative samples, on the other hand, are still generated on the GPU: the kernel for the parts $(V_{i}^{j}, V_{i}^{k})$ draws the negative samples for vertices in $V_{i}^{j}$ randomly from $V_{i}^{k}$ and vice versa for vertices in $V_{i}^{k}$.\looseness=-1

\subsubsection{Minimizing the data movements:}
Since the embedding is performed on pairs of sub-matrices, the rotation order in which we process the pairs is important to minimize the data movement operations. We follow an order resembling the {\em inside-out order} proposed in~\cite{pbg19} which
formally defines the part pairs as $(V_i^{a_{0}}, V_i^{b_{0}}), (V_i^{a_{1}}, V_i^{b_{1}}), \cdots, (V_i^{a_{\ell}}, V_i^{b_{\ell}})$ where $\ell = \frac{K_{i}(K_{i}+1)}{2}$ and
\[(V_i^{a_j}, V_i^{b_j}) = 
\begin{cases}
(V_i^{0}, V_i^{0}) & $j = 0$ \\\
(V_i^{a_{j-1}}, V_i^{b_{j-1}+1}) & j > 0 \text{ and } a_{j-1} > b_{j-1}  \\
(V_i^{a_{j-1}+1}, V_i^{0}) & a_{j-1} = b_{j-1} \\
\end{cases}\]

\subsubsection{Choosing the sub-matrices and samples to be stored:} 
Let $P_{GPU}$ be the number of sub-matrices stored on the GPU at a time. Since we require every sub-matrix pair to exist on the GPU together during a single rotation, the smallest acceptable value is $2$. However, $P_{GPU} = 2$  means that there will be time instances where all the kernels processing the current sub-matrices finish and a new kernel cannot start until a new sub-matrix is copied to the GPU. This leaves the GPU idle during the copy operation. On the other hand, using $P_{GPU} > 2$ occupies more space but allows an overlap of data transfers with kernel executions. For instance, assume ${\mathbf M}_i^1, {\mathbf M}_i^2$ and ${\mathbf M}_i^4$ are on GPU and the three upcoming kernels are $(V_i^4, V_i^1), (V_i^4,V_i^2)$ and $(V_i^4,V_i^3)$. The first two kernels are dispatched and after the first finishes, while the second is running, ${\mathbf M}_i^1$ is replaced with ${\mathbf M}_i^3$, thus hiding the latency. 
A large $P_{GPU}$ increases the amount of overlap. However, it also consumes more space on the GPU and increases $K_{i}$, i.e., the number of sets in ${\mathcal V}_i$. This leads to a rotation containing more kernels, i.e., pairs to be processed. 
For this reason, we set $P_{GPU} = 3$ to both hide the latency and occupy less GPU memory.\looseness=-1

Since we do not keep the large graphs on GPU memory and draw positive samples on the CPU, these samples must also be transferred to the GPU for each kernel. However, if all these samples are transferred at once, similar to above, $K_i$ increases and the performance decreases. To solve this issue, we only keep samples for $S_{GPU}$ pairs on the GPU and dynamically replace a pool once it is consumed. We set $S_{GPU} = 4$ as we've experimentally found it to be an adequate value for all our graphs. 

\newcommand{\samplemanager}{{\tt SampleManager}\xspace}
\newcommand{\poolmanager}{{\tt PoolManager}\xspace}
\newcommand{\switchsubmatrices}{{\sc SwitchSubMatrices}\xspace}
\newcommand{\embeddingkernel}{{\sc EmbeddingKernel}\xspace}
\newcommand{\getnextsubmatrix}{{\sc NextSubMatrix}\xspace}
\subsubsection{Implementation details:}
Embedding a large graph $G_{i}$ requires an orchestrated execution of multiple tasks on the host and the device. \malgo coordinates the following tasks by single host thread:\looseness=-1 
\begin{enumerate}
\item The main thread dispatches the embedding kernels to the GPU, as well as moves the sub-matrices forth and back between the host and GPU. Multiple GPU streams are used to allow for multiple kernel dispatches at once to maximize the utilization.
\item The \samplemanager thread performs (positive) sampling into pools. When necessary, \samplemanager creates a team of threads to generate samples for a single sample pool. Once a pool is 
ready to be sent to the GPU, it is kept in a buffer.
\item \poolmanager dispatches the sample pools to the GPU. Once a sample pool is used up on the GPU, and becomes free to be overwritten, \poolmanager dispatches a ready pool to the GPU. 
\end{enumerate}

\begin{figure}
    \centering
    \includegraphics[width=\linewidth]{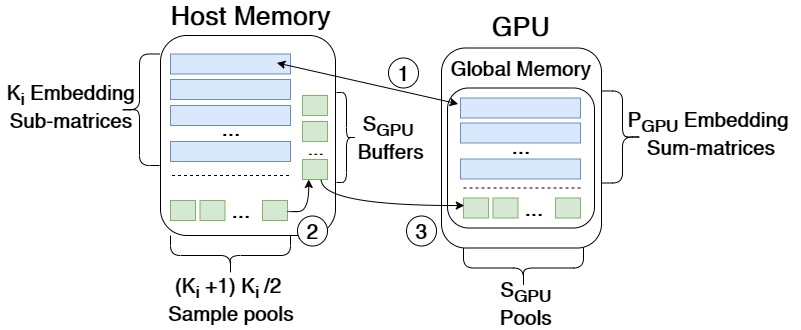}
    \caption{Memory model of large graphs algorithm embedding graph $G_{i}$. 1) Embedding sub-matrices are copied between the host and GPU as needed, 2) When sample pool $S_{i}^{j, k}$ is ready, it is copied to an empty buffer. 3) When a sample pool on the GPU is used up, it is replaced by the next sample pool from the buffer.}
    \label{fig:large_graphics_memory}
    \Description[Fully described in the text]{Fully described in the text}
\end{figure}

Coordination among these threads is provided through condition variables. 
A high-level overview of the large graph embedding process is shown in Figure~\ref{fig:large_graphics_memory} and its pseudocode is given in Algorithm~\ref{alg:LGE}.\looseness=-1

In Line~\ref{ln:lge:K} of Algorithm~\ref{alg:LGE}, we compute $K_{i}$ and the number of rotations $e`$. Line~\ref{ln:lge:gstate} initializes an array $GPUState$ of size $P_{GPU}$ which keeps track of embedding sub-matrices currently stored on the GPU. $GPUState[j] = k$ means that bin $j$ on the GPU currently holds $M_{i}^k$ where an entry $-1$ indicates that the bin $j$ is empty. At the beginning, the first $P_{GPU}$ sub-matrices are copied to the GPU~(lines~\ref{ln:lge:for}--\ref{ln:lge:init}) by calling the function \switchsubmatrices$(j, k)$ which copies ${\mathbf M}_{i}^{j}$ out of the GPU, replaces it with ${\mathbf M}_{i}^{k}$, and returns the new $GPUState$. Afterwards, the \poolmanager and \samplemanager threads are started and the main thread starts to perform the embedding rotations. Lines~\ref{ln:lge:mainfor}--\ref{ln:lge:mainforend} run the main embedding loop. 

When the sub-matrices ${\mathbf M}_{i}^{m}$, ${\mathbf M}_{i}^{s}$ and the sample pool $S_{i}^{m, s}$ are ready, the function \embeddingkernel(${\mathbf M}_{i}^{m}$, ${\mathbf M}_{i}^{s}$, $n_s$, $lr$) runs the embedding kernel on the sub-matrices pair $({\mathbf M}_{i}^{j}, {\mathbf M}_{j}^{k})$ using $n_s$ negative samples and learning rate $lr$. Finally, the function \getnextsubmatrix($GPUState$, $a$, $b$) will determine the next sub-matrix to be switched into the GPU after running \embeddingkernel(${\mathbf M}_{i}^{j}$, ${\mathbf M}_{i}^{k}$, $n_s$, $lr$) given $GPUState$.

\begin{algorithm}[htbp]
\DontPrintSemicolon
\LinesNumbered
\KwData{$G_{i}$, ${\mathbf M}_{i}$, $n_{s}$, $lr$, $e_{i}$, $P_{GPU}$, $S_{GPU}$, $B$}
\KwResult{${\mathbf M}_{i}$}
    $e' \leftarrow \frac{e_{i}}{B\text{{$\times K_{i}$}}}$, $K_{i} \leftarrow$  {\sc GetEmbeddingPartInfo}($G_{i}$) \;\label{ln:lge:K}
    $GPUState$[$0:P_{GPU}-1$] $\leftarrow$ \{$-1$, $-1$, $\dots$, $-1$\}\; \label{ln:lge:gstate}
    \For{$i \leftarrow  0$ to $P_{GPU} - 1$}{  \label{ln:lge:for}
        $GPUState$ $\leftarrow$  \switchsubmatrices(-1, $i$)\; \label{ln:lge:init}
    }
    {\sc thread}({\tt SampleManager}, $K_{i}$, $e'$, $S_{GPU}$)  \;
    {\sc thread}({\tt PoolManager}, $K_{i}$, $e'$, $S_{GPU}$) \;
    
    \For{$r$ from $1$ to $e'$}{ \label{ln:lge:mainfor}
        \For{$m$ from $0$ to $K_{i} - 1$}{
            \For{$s$ from $0$ to $m$}{
                Wait for ${\mathbf M}^m_{i}$, ${\mathbf M}^s_{i}$ and $S_{i}^{m, s}$ to be on GPU\;
                \embeddingkernel(${\mathbf M}_{i}^{m}$, ${\mathbf M}_{i}^{s}$ ,$n_{s}$, $lr$)\;
                 $nextSM \leftarrow$  \getnextsubmatrix($GPUState$, $m$, $s$)\;
                 $GPUState$ $\leftarrow$  \switchsubmatrices($s$, $nextSM$)\;
            }
            
        }\label{ln:lge:mainforend}
    }
\caption{\lge}\label{alg:LGE}
\end{algorithm}
\section{Experiments}\label{sec:exp}

We will first explain the ML pipeline we used to evaluate embeddings and to compute AUCROC~(Area Under the Receiver Operating Characteristics) scores. Then the data-sets will be summarized and the state-of-the-art tools used to evaluate the performance of \malgo will be listed. Lastly, the results will be given. 

\subsection{Evaluation with link-prediction pipeline}\label{sec:pipeline}
We evaluate the embedding quality of \malgo, \versex, \mile, and \graphvite with link prediction, which is one of the most common ML tasks that  embedding algorithms are evaluated by~\cite{pbg19, graphvite19, verse18, node2vec}.
First, the input graph $G$ is split into train and test sub-graphs as $G_{train} = (V_{train}, E_{train})$ and $G_{test} = (V_{test}, E_{test})$ respectively. $G_{train}$ contains 80$\%$ of the edges of $G$, where $G_{test}$ contains the remaining 20$\%$. Then, we remove all the isolated vertices from $G_{train}$ and also all $(u,v)$ edges from $G_{test}$, where $u$ or $v$ is not in $G_{train}$.
This guarantees that $V_{test} \subseteq V_{train}$.
Next, we execute the tools to generate embeddings of $G_{train}$.
Finally, we employ a Logistic Regression model which uses the embeddings generated in the previous step to predict the existence of edges in $G_{test}$. For medium scale graphs, we used the {\tt LogisticRegression} module from {\tt scikit-learn}. However, logistic regression becomes too expensive for large-scale graphs. Thus, for such graphs, we use the {\tt SGDClassifier} module from {\tt scikit-learn} with a logistic regression solver. 

For the prediction pipeline, we create two matrices, ${\mathbf R}_{train}$, and ${\mathbf R}_{test}$. 
Each vector ${\mathbf R}_{train}[i]$ represents either a positive or a negative sample, and we obtain these vectors by doing element-wise multiplication of two vectors from ${\mathbf M}_0$, corresponding to two vertices in the graph.  
${\mathbf R}_{train}$ includes all the edges in  $G_{train}$ as positive samples. 
Moreover, we generate $|E_{train}|$ number of negative samples from $(V_{train} \times V_{train}) \setminus E_{train}$ and add them as vectors to ${\mathbf R}_{train}$ to make a balanced training set for the logistic regression classifier. 
In addition to $d$ values obtained via element-wise multiplications, for ${\mathbf R}_{train}$, a label representing a positive or negative sample is concatenated to the end of the vector. Hence, the length of each ${\mathbf R}_{train}$ vector is $d+1$. 
We create ${\mathbf R}_{test}$ in a similar fashion by using $G_{test}$ instead of $G_{train}$ as the source of the samples. We first train the logistic regression model using ${\mathbf R}_{train}$, and then test the validity of the model with ${\mathbf R}_{test}$. Finally, we report the \emph{AUCROC} score obtained from the test set~\cite{roc}.\looseness=-1


\subsection{Datasets used for the experiments}
We use various graphs in the evaluation process to cover many structural variations and to evaluate the tools in terms of performance and quality as fairly and thoroughly as possible. The graphs differ in terms of their origin, the number of vertices, and density. The properties of these graphs are given in Table \ref{table:graph_summary}. The medium-scale graphs, with less than 10M vertices, and large-scale ones are separated in the table.

\begin{table}
\caption{Normal and large graphs used in the experiments.}
\label{table:graph_summary}
\scalebox{0.85}{
\begin{tabular}{lrrr}
\textbf{Graph}  & \textbf{|V|} & \textbf{|E|} & \textbf{Density} \\
\midrule
{\tt com-dblp}~\cite{snapnets}        & 317,080      & 1,049,866      & 3.31             \\ 
{\tt com-amazon}~\cite{snapnets}       & 334,863      & 925,872      & 2.76             \\ 
{\tt youtube}~\cite{yt}         & 1,138,499    & 4,945,382    & 4.34             \\ 
{\tt soc-pokec}~\cite{snapnets}        & 1,632,803 & 30,622,564   &    18.75         \\ 
{\tt wiki-topcats}~\cite{snapnets}     & 1,791,489  &  28,511,807  &     15.92        \\ 
{\tt com-orkut}~\cite{snapnets}        & 3,072,441    & 117,185,083   &  38.14           \\ 
{\tt com-lj}~\cite{snapnets}           & 3,997,962     &  34,681,189  &  8.67            \\ 
{\tt soc-LiveJournal}~\cite{snapnets}  & 4,847,571    &  68,993,773  & 14.23  \\\hline
{\tt hyperlink2012}~\cite{hl}   & 39,497,204    & 623,056,313       & 15.77            \\ 
{\tt soc-sinaweibo}~\cite{nr}   &  58,655,849   &   261,321,071     &  4.46            \\ 
{\tt twitter\_{rv}}~\cite{nr}      & 41,652,230    &  1,468,365,182    & 35.25             \\ 
{\tt com-friendster}~\cite{snapnets}   & 65,608,366    &   1,806,067,135   &   27.53           \\ 
\end{tabular}
}
\end{table}

\subsection{State-of-the art tools used for evaluation}\label{comp_sys}
To evaluate the performance and the quality of \malgo, we use the results of the following state-of-the-art tools as a baseline.

\versex: is a recent multi-core graph embedding tool~\cite{verse18}. It can employ different vertex-similarity measures including PPR, adjacency lists, and SimRank. We use the PPR similarity measure and $\alpha = 0.85$ as recommended by the authors. For \versex, we set the epoch number to $e = 600, 1000$, and $1400$, use a learning rate of $lr = 0.0025$ and report the best AUCROC. Larger learning rates produce worse results.

\graphvite: is a state-of-the-art, fast multi-GPU graph embedding tool. However, according to~\cite{graphvite19}, the algorithm cannot embed graphs with $|V| > 12,000,000$ on a single GPU. We use the default values for the hyperparameters as recommended by the authors and LINE is chosen as the base embedding method. As for the number of epochs, we use two settings; a fast setting with $e = 600$ epochs, and a slow setting with $e = 1000$ epochs.

\mile: performs embedding by coarsening a graph into multiple levels similar to \malgo~\cite{mile18}. It trains the smallest level and refines the embeddings up the coarsening levels. However, unlike \malgo, \mile uses a neural network to project the coarsened graph embedding to that of the original graph. We use the following parameters for the model: {\sc DeepWalk} as a base embedding method, {\sc MD-GCN} as a refinement method, 8 levels of coarsening, and a learning rate of $lr = 0.001$. As for the epochs used during training, we note that \mile does not allow the number of epochs to be configured. Hence, that decision was left to the model itself.

For \malgo, we use three configurations: fast, normal and slow, with parameters given in Table~\ref{table:configs}. The configurations differ in terms of the number of epochs, smoothing ratio and learning rate. Compared to \malgo-slow, \malgo-fast uses a smaller $p$ and a larger $lr$ to compensate for the lesser number of epochs on the original graph with faster learning. Furthermore, we include in the experiments a version of \malgo which does not perform coarsening. This configuration spends all of the epochs on the original graph. In addition to these differences, for medium-scale graphs, a larger number of epochs is used for each configuration compared to large-scale graphs. 

For the experiments with \malgo and \versex, we define a single epoch as sampling $|E|$ target vertices. We do so to match the definition of an epoch given by \graphvite~\cite{graphvite19} for the fairness of the experiments. With this definition, using fewer epochs for large-scale graphs makes more sense since a single epoch implies billions of samples and updates for such graphs, whereas for medium-scale graphs this number is in the order of tens of millions.

\begin{table}
\caption{\malgo configurations, fast, normal and small for medium-scale and large-scale graphs. A version with no coarsening is also used in the experiments. }
\label{table:configs}
\scalebox{0.88}{
\begin{tabular}{l|rr|r|r}
\textbf{Configuration} & $p$ & $lr$ & $e_{normal}$ & $e_{large}$\\
\midrule
Fast   & 0.1 & 0.050 & 600 & 100 \\
Normal & 0.3 & 0.035 & 1000 & 200 \\
Slow   & 0.5 & 0.025 & 1400 & 300 \\\hline
No coarsening & - & 0.045 & 1000 & 200  \\
\end{tabular}
}
\end{table}

All the experiments run on a single machine with 2 sockets, each with 8 Intel
E5-2620 v4 CPU cores running at 2.10GHz with two hyper-threads per core~(32 logical cores in total), and 198GB RAM. To avoid the effects of hyper-threading, we only use 16 threads for parallel executions. The GPU experiments use a single Titan X Pascal GPU with 12GB of memory. 
The server has {\tt Ubuntu 4.4.0-159} as the operating system. All the CPU codes are compiled with {\tt gcc 7.3.0} with {\tt -O3} as the optimization parameter. For CPU parallelization, {\tt OpenMP} multithreading is used in general. Only for large graphs, a hybrid implementation with {\tt OpenMP} and {\tt C++11} threading is employed. For GPU implementations and compilation, {\tt nvcc} with {\tt CUDA 10.1} and optimization flag {\tt -O3} are used. The GPUs are connected to the server via {\tt PCIe 3.0 x16}. For GPU implementations, all the relevant data structures are stored on the device memory, unified memory is not used.

\subsection{Experiments on coarsening performance} \label{coarsening-performance}


Table~\ref{table:seq_parallel_coarsening} provides the properties of the coarsenings obtained from the sequential and parallel coarsening algorithms with $\tau = 32$ threads. 
As the results show, parallel coarsening reaches a similar number of levels and the graphs at the last-level are of similar sizes. 
Hence, there is a negligible difference regarding the quality of graphs generated by the two algorithms. Only for {\tt soc-sinaweibo}, there is a one-level difference for which $|V_{D - 1}|$ also has a difference of 142 vertices.\looseness=-1

With a similar coarsening quality, the parallel algorithm is $5$--$10\times$ faster compared to the sequential counterpart. As described in Section~\ref{s_coarsening_parallel}, the 
time complexity is ${\mathcal O}$($|V| + |E|$) and in practice, $|E|$ dominates the workload. Although there are other parameters, the variation in the speedups is in concordance with the variation in the number of edges. For instance, {\tt soc-sinaweibo} only has $200$M edges and yields the smallest speedup value of $5.8\times$. On the other hand, the largest speedup $10.5\times$ is obtained for {\tt com-friendster}, which is the largest in our data-set with $1.8$B edges.\looseness=-1 

\begin{table}[htbp]
\caption{Execution times, the number of levels and the size of the last-level graphs for sequential and parallel coarsening with $\tau = 32$ threads for the large-scale graphs. The results are the average of 5 runs.}
\label{table:seq_parallel_coarsening}
\scalebox{0.88} {
\begin{tabular}{lrrrrr}
\textbf{Graph}  & $\tau$ &\textbf{Time (s)} & Speedup & $D$ & \textbf{$|V_{D - 1}|$}\\
\midrule
\multirow{2}{*}{{\tt  hyperlink2012}} & 1  & 365.49  & -  & 8       & 2411            \\
& 32 & 45.36  & 8.06$\times$  &   8  & 2385          \\\cline{2-6}
\multirow{2}{*}{{\tt soc-sinaweibo}} & 1   & 135.92 & -    & 10       & 272             \\
& 32  & 23.54   &  5.77$\times$    & 9   & 414\\\cline{2-6}
\multirow{2}{*}{{\tt twitter\_rv}} & 1      & 629.20 & -    & 12     & 541             \\
& 32 & 77.77  &  8.09$\times$  & 12 & 432  \\\cline{2-6}
\multirow{2}{*}{{\tt com-friendster}} & 1  & 2468.52  & -   & 10     & 1164             \\
& 32 & 235.38  &  10.49$\times$  & 10 & 1158      \\
\end{tabular}
}
\end{table}

\subsubsection{\textbf{\malgo} vs \mile}

In Table \ref{table:MILEvsO}, we show a brief comparison of \emph{MILE} and \malgo with $16$ threads on the graph {\tt com-orkut}. Since \emph{MILE} does not have a stopping criterion for coarsening, we used the same amount of coarsening levels for both algorithms. While coarsening a graph of 3 million vertices and 100 million edges, \malgo is 264 times faster than \emph{MILE}. Moreover, \malgo is a lot more efficient regarding the number of vertices obtained at each level. For instance, in 8 levels \malgo shrinks {\tt com-orkut} to only 230 vertices, while \emph{MILE} shrinks it to 12062 vertices. This is important since the training time is affected by the number of vertices at each level.\looseness=-1

\begin{table}
\caption{\mile vs \malgo coarsening on {\tt com-orkut}. A parallel coarsening with $\tau = 16$ threads is used for \malgo.}
\label{table:MILEvsO}
\scalebox{0.88}{
\begin{tabular}{lrrr|lrrr}
\textbf{} & $i$ & \textbf{Time (s)} & \textbf{$|V_{i}|$} & \textbf{} & $i$ & \textbf{Time (s)} & \textbf{$|V_{i}|$} \\
\midrule
\multirow{9}{*}{\rotatebox[origin=c]{90}{\mile}} & 0 & - & 3056838 & \multirow{9}{*}{\rotatebox[origin=c]{90}{\malgo}} & 0 & - & 3056838 \\
            &1 & 249.77 & 1535168 && 1 & 4.44 & 975132 \\ 
            &2 & 237.39 & 768804 && 2 & 1.23 & 213707 \\
            &3 & 184.72 & 384752 && 3 & 0.62 & 46667 \\
            &4 & 151.24 & 192507 && 4 & 0.16 & 8084 \\
            &5 & 139.23 & 96308 && 5 & 0.03 & 2000 \\
            &6 & 128.47 & 48183 && 6 & 0.01 & 701 \\
            &7 & 117.75 & 24107 && 7 & < 0.01 & 375 \\
            &8 & 99.73 & 12062 && 8 &  < 0.01  & 275 \\\hline
            &Total & 1308.31 & - && Total & 6.60 & - \\
\end{tabular}
}
\end{table}

\begin{table*}[h!]
  \caption{Link prediction results on medium-scale graphs. Every data-point is the average of 15 results. \versex and \malgo uses $\tau = 16$ threads. \mile is a sequential tool. Both \graphvite and \malgo uses the same GPU. The speedup values are computed based on the execution time of \versex.}
  \label{tab:medium_results}
  \scalebox{0.82}{
\begin{tabular}{llrrr||llrrr}
     {\bf Graph} & {\bf Algorithm} & {\bf Time (s)} & {\bf Speedup} & {\bf AUCROC}(\%) & {\bf Graph} & {\bf Algorithm} & {\bf Time (s)} & {\bf Speedup} & {\bf AUCROC}(\%)  \\
     \midrule
     \multirow{7}{5em}{{\tt com-dblp}} & \versex & 247.99 & 1.00$\times$ & {\bf 97.82} & \multirow{7}{5em}{{\tt com-amazon}} & \versex & 216.18 &$1.00\times$ & 97.71 \\\cline{2-5}\cline{7-10}
          & \mile &  136.65 & 1.81$\times$ & 97.65 & & \mile & 146.29 & 1.48$\times$ & {\bf 98.14} \\\cline{2-5}\cline{7-10}
     & \graphvite-fast & 13.97 & 17.70$\times$ & 97.80 & & \graphvite-fast & 12.45 & 17.36$\times$ & 97.40 \\
     & \graphvite-slow & 19.93 & 12.40$\times$ & \textbf{98.08} & & \graphvite-slow & 16.84 & 12.83$\times$ & 97.82 \\\cline{2-5}\cline{7-10}
     & \malgo-fast & 0.72  & 344.43$\times$& 96.45 & & \malgo-fast & 0.69 & 313.30$\times$ & 97.20 \\
     & \malgo-normal & 2.08  & 119.23$\times$& 97.38 & & \malgo-normal & 1.88 & 114.99$\times$ & 98.29 \\
     & \malgo-slow & 3.84  & 64.58$\times$& 97.63 & & \malgo-slow & 3.59 & 60.22$\times$ & \textbf{98.43} \\\cline{2-5}\cline{7-10}
     & \malgo-NoCoarse & 29.97 & 8.27$\times$ & 93.31 & & \malgo-NoCoarse & 24.60 & 8.79$\times$ & 90.13 \\
     \hline
     \multirow{6}{5em}{{\tt com-lj}} & \versex & 12502.72 & 1.00$\times$ & \textbf{98.86} & \multirow{6}{5em}{{\tt com-orkut}} & \versex & 45994.93 &1.00$\times$  &  \textbf{98.65} \\\cline{2-5}\cline{7-10}
     & \mile & 3948.62 & 3.17$\times$ & 80.19 & & \mile & 11904.31 & 3.86$\times$ & 90.38 \\\cline{2-5}\cline{7-10}
     & \graphvite-fast & 373.58 & 33.47$\times$ & 98.04 & & \graphvite-fast & 1246.38 & 36.90$\times$ & 98.02 \\
     & \graphvite-slow & 644.43 & 19.40$\times$ & 98.33 & & \graphvite-slow & 2199.25 & 20.91$\times$ & {\bf 98.05} \\\cline{2-5}\cline{7-10}
     & \malgo-fast & 16.27 & 768.45$\times$ & 96.82 & & \malgo-fast & 43.30 & 1062.24$\times$ & 97.35 \\
     & \malgo-normal & 55.01 & 227.28$\times$ & 98.33 & & \malgo-normal & 185.12 & 248.46$\times$ & 97.63 \\
     & \malgo-slow & 153.72 & 81.33$\times$ & {\bf 98.46} & & \malgo-slow & 487.33 & 94.38$\times$ & 97.69 \\\cline{2-5}\cline{7-10}
     & \malgo-NoCoarse & 675.25 & 18.52$\times$ & 98.32 & & \malgo-NoCoarse & 2301.89 & 19.98$\times$ & 97.64 \\
     \hline
     \multirow{6}{5em}{{\tt wiki-\\topcats}} & \versex & 8709.48 & 1.00$\times$ & \textbf{99.31} & \multirow{6}{5em}{{\tt youtube}} & \versex & 1365.36 & 1.00$\times$ & \textbf{98.04}\\\cline{2-5}\cline{7-10}
     & \mile & 4953.68 & 1.76$\times$ & 86.04 & & \mile & 1328.62 & 1.03$\times$ & 94.17 \\\cline{2-5}\cline{7-10}
     & \graphvite-fast & 310.47 & 28.05$\times$ & 96.42 & & \graphvite-fast & 63.90 & 21.37$\times$ & 97.07 \\
     & \graphvite-slow & 544.06 & 16.01$\times$ & 96.28 & & \graphvite-slow & 104.76 & 13.03$\times$ & 97.45 \\\cline{2-5}\cline{7-10}
     & \malgo-fast & 11.34 & 768.03$\times$ & 98.13 & & \malgo-fast & 2.76 & 494.70$\times$ & 96.16 \\
     & \malgo-normal & 40.76 & 213.68$\times$ & 98.33 & & \malgo-normal & 7.15 & 190.96$\times$ & 97.78 \\
     & \malgo-slow & 93.86 & 92.79$\times$ & {\bf 98.50} &  & \malgo-slow & 15.32 & 89.12$\times$ & {\bf 97.93} \\\cline{2-5}\cline{7-10}
     & \malgo-NoCoarse & 549.65 & 15.85$\times$ & 98.51 & & \malgo-NoCoarse & 158.60 & 8.61$\times$ & 97.16 \\
     \hline
     \multirow{6}{5em}{{\tt soc-pokec}} & \versex & 9182.53 & 1.00$\times$ & \textbf{98.32} & \multirow{6}{5em}{{\tt soc-\\LiveJournal}} & \versex & 14965.76 & 1.00$\times$ &  {\bf 97.61} \\\cline{2-5}\cline{7-10}
     & \mile & 2848.78 & 3.22$\times$ & 85.75 & & \mile & 6210.58 & 2.41$\times$ & 80.84 \\\cline{2-5}\cline{7-10}
     & \graphvite-fast & 370.73 & 24.77$\times$ & {\bf 97.42} & & \graphvite-fast & 745.33 & 20.08$\times$ & 99.23 \\
     & \graphvite-slow & 607.07 & 15.13$\times$ & 97.37 & & \graphvite-slow & 1209.95 & 12.37$\times$ & \textbf{99.31} \\\cline{2-5}\cline{7-10}
     & \malgo-fast & 16.34 & 561.97$\times$ & 96.34 & & \malgo-fast & 29.74 & 503.22$\times$ & 98.58 \\
     & \malgo-normal & 54.66 & 167.99$\times$ & 96.49 & & \malgo-normal & 112.72 & 132.77$\times$ & 98.87 \\
     & \malgo-slow & 131.06 & 70.06$\times$ & 96.67 & & \malgo-slow & 183.64 & 81.50$\times$ & 98.76 \\\cline{2-5}\cline{7-10}
     & \malgo-NoCoarse & 598.95 & 15.33$\times$ & 97.28 & & \malgo-NoCoarse & 1348.74 & 11.10$\times$ & 98.88 \\
     \bottomrule
\end{tabular}
}
\end{table*}

\subsection{Experiments on handling large graphs}\label{lges}

\begin{figure}
    \centering
    \includegraphics[width=0.90\linewidth]{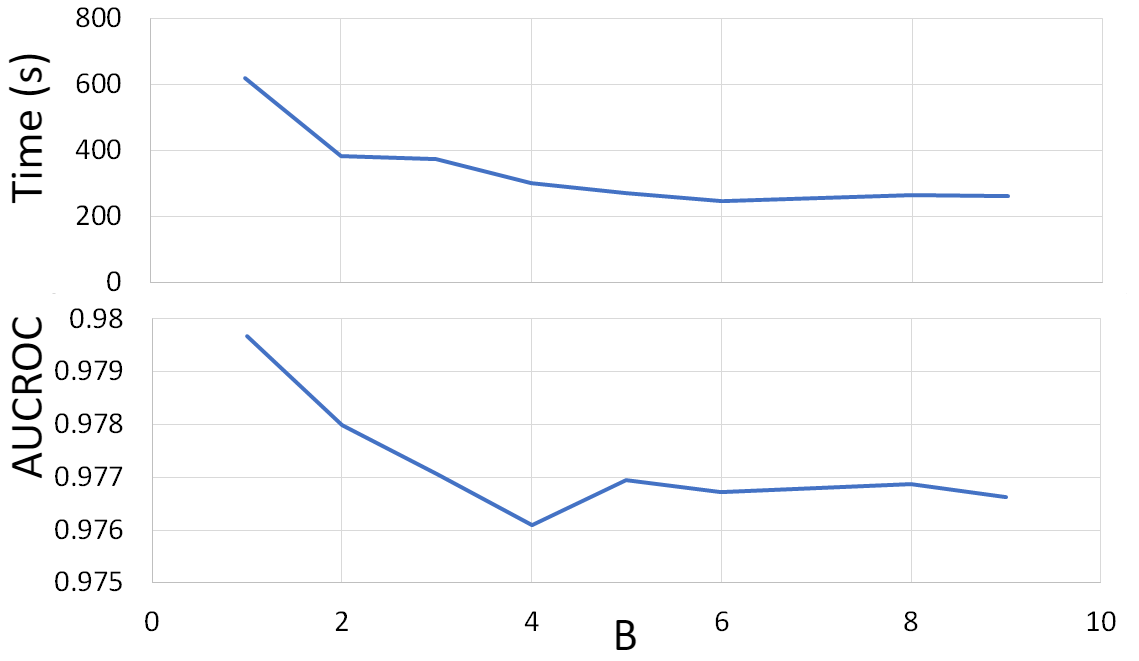}
    \caption{Running large-graph embedding on {\tt hyperlink} with different $B$ values.}
    \label{fig:bvauctime}
    \Description{As the value of B increases, the embedding runtime decreases but so does the AUCROC achieved from the embedding}
\end{figure}
Figure~\ref{fig:bvauctime} shows the effect of adjusting the batch size $B$ for large-graph embedding. The top figure provides the execution time and the bottom figure provides the AUCROC scores. We can see the trade-off between performance and quality while increasing $B$. The execution time decreases since the number of embedding rounds is reduced. 
However, the quality also decreases since 
increasing $B$ results in increasing the number of updates being carried out on a subset of the graph {\em in isolation} from the rest of the embedding process. 
To be as efficient as possible and not to decrease the accuracy significantly, we use $B = 5$ as the default value for \malgo.

\subsection{Experiments on embedding quality}

Tables~\ref{tab:medium_results} and~\ref{tab:large_results} provide the execution times and AUCROC scores of the tools evaluated in this work on medium-scale and large-scale graphs, respectively. The parameters for the tools are given in Section~\ref{comp_sys}~(see Table~\ref{table:configs} for \malgo configurations).
The first observation is that coarsening does not have a significant negative effect on the quality of the embedding. While \malgo performs worse on some medium-scale graphs with coarsening, for others, the scores are approximately the same or even better. Since all of the epochs are reserved for the original graph in the non-coarsened version, we expect the results of this configuration to be more fine-tuned. However, the data shows that training in the coarser levels may have a more prominent effect than training more on the original graph.\looseness=-1

In addition to \malgo configurations, we use \graphvite, \mile, and \versex on medium-scale graphs. The results are given in Table~\ref{tab:medium_results}.
To evaluate the runtime performance of the tools, we use the execution time of \versex as the baseline and present the speedups by each tool. From these experiments, we have the following observations:
\begin{itemize}
    \item \malgo-fast is an ultra-fast solution that produces very accurate embeddings at a fraction of the time compared to all the systems under evaluation. It can achieve a speedup over \versex of up to three orders of magnitude and an average of $600\times$ with a maximum loss in AUCROC of $2\%$ and an average loss of $1.16\%$. When compared to \mile, it is superior in terms of AUCROC in six out of eight graphs while being at least two orders of magnitude faster. As for \graphvite, we see that, at an average loss in AUCROC of $0.54\%$, \graphvite can achieve an average speedup of $23.44\times$. 
    \item \malgo-normal demonstrates the speed/quality trade-off of \malgo and its flexibility. Switching from \malgo-fast to \malgo-normal results in an average AUCROC increase of $0.76\%$ while only reducing the speed on average by a factor of $3\times$.\looseness=-1
    \item The flexibility is demonstrated further by \malgo-slow whose accuracy becomes close to the best tool for every graph. Compared to \versex, this configuration has an average loss of $0.24\%$ in AUCROC, but still has an average speedup of $79.24\times$.\looseness=-1   
    \item To compare \malgo with the state-of-the-art GPU implementation \graphvite, we used the best AUCROCs of the tools. For 4/8 graphs, \malgo configurations produce better AUCROC compared to \graphvite configurations. The values are similar; on average, \malgo achieves $0.16\%$ higher AUCROCs than \graphvite. However, \malgo is $5.2\times$ faster than \graphvite on average.
\end{itemize}

\subsubsection{Large-scale graphs}
The results of the experiments for large-scale graphs can be seen in Table~\ref{tab:large_results}. We observe the following:

\graphvite results are not reported since, for all the large-scale graphs, the executable runs out of CPU memory on our machine. We find that \graphvite on {\tt hyperlink2012} is reported to achieve $94.3\%$ link-prediction AUCROC after an embedding for 5.36 hours using four Tesla P100s GPUs~\cite{graphvite19}. \malgo-normal achieves an AUCROC of $97.20\%$ after an embedding taking only 0.2 hours using a single Titan X GPU~($26.8\times$ speedup). It is also reported that \graphvite takes 20.3 hours on {\tt com-friendster}~\cite{graphvite19} where \malgo-normal requires only 0.76 hours~($26.7\times$ speedup).

\mile cannot embed {\tt hyperlink2012} and {\tt soc-sinaweibo} before the 12 hour timeout. Furthermore, the executable failed to run on the other two graphs due to insufficient memory.

\versex timed out for 3 of the 4 graphs, as shown in Table~\ref{tab:large_results}. However, it performed the embedding on {\tt soc-sinaweibo}  successfully. Compared to \malgo-slow, it scores a $0.52\%$ higher AUCROC. On the other hand, \malgo-slow achieves a $26\times$ speedup over \versex. 

\begin{table}[htbp]
  \caption{Link prediction results on large graphs. Every data-point is the average of 6 results. \graphvite and \mile fail to embed any of the graphs due to excessive memory usage or an execution time larger than 12 hours. $\tau = 16$ threads used for both \versex and \malgo.}
  \label{tab:large_results}
  \scalebox{0.88}{
\begin{tabular}{llrrr}
     &&&&{\bf AUC}\hspace*{3.3ex} \\
     {\bf Graph} & {\bf Algorithm} & {\bf Time (s)} & {\bf Speedup} & {\bf ROC} (\%)  \\
     \toprule
     \multirow{4}{*}{{\small{{\tt hyperlink2012}}}} & \versex & Timeout & - &- \\\cline{2-5}
     & \malgo-fast & 201.02 & - & 87.60  \\
     & \malgo-normal & 724.09 & - & 97.20  \\
     & \malgo-slow & 1676.93 & -  & 98.00  \\
     \toprule
     \multirow{4}{*}{{\small{{\tt soc-sinaweibo}}}}  & \versex & 20397.79 & 1.00$\times$ & 99.89 \\ \cline{2-5}
     & \malgo-fast & 48.88 & 417.30$\times$ & 70.27  \\
     & \malgo-normal & 352.86 & 57.81$\times$ & 97.00  \\
     & \malgo-slow & 759.85 & 26.84$\times$ & 99.37 \\
     \toprule
     \multirow{4}{*}{{\small{{\tt twitter\_rv}}}} & \versex & Timeout & -& - \\\cline{2-5}
     & \malgo-fast & 261.08 & -  & 91.78 \\
     & \malgo-normal & 994.46 & -  & 97.36 \\
     & \malgo-slow & 2128.70 & -  & 98.50  \\
     \toprule
     \multirow{4}{*}{{\small{{\tt com-friendster}}}} & \versex & Timeout & - & -\\\cline{2-5}
     & \malgo-fast & 680.33 & -  & 85.17  \\
     & \malgo-normal & 2720.82 & -  & 93.40 \\
     & \malgo-slow & 5000.96 & -  & 94.98\\
\end{tabular}
}
\end{table}

\subsection{Experiments on smaller dimensions}

We analyzed the performance of \malgo when multiple vertices are assigned to a single warp, where $d$ is small. 
The results on {\tt com-orkut} and {\tt soc-LiveJournal} are given in Table~\ref{table:small_dimensions}.
Without small-dimension technique~(SM), \malgo takes approximately the same time for $d = 8, 16$ and $32$ where $4\times$ and $2\times$ less work is performed for $d = 8$ and $16$. With SM, we observe $2.63\times$ and $1.84\times$ speedups for $d = 8$ and $16$, respectively. Moreover, for {\tt soc-LiveJournal}, we obtain $2.70\times$ and $1.85\times$ speedups for $d = 8$ and $d = 16$, respectively. As expected, with or without SM, $d = 32$ timings are almost the same.\looseness=-1 

\begin{table}[h!]
\caption{Performance of \malgo with~(SM = Yes) \& without~(SM = No) small-dimension embedding and $\tau = 16$ threads.}
\label{table:small_dimensions}
\scalebox{0.88}{
\begin{tabular}{ccrr|ccrr}
 \textbf{Graph} & \textbf{SM}& \textbf{$d$} & \textbf{Time (s)} & \textbf{Graph} & \textbf{SM}& \textbf{$d$} & \textbf{Time (s)}\\
\midrule
    \multirow{6}{*}{\rotatebox[origin=c]{90}{{\tt com-orkut}}}  & \multirow{3}{*}{No} & 8 & 63.72 & \multirow{6}{*}{\rotatebox[origin=c]{90}{{\small{\tt soc-LiveJournal}}}}  & \multirow{3}{*}{No} & 8 & 40.13 \\
    & & 16 & 64.20 & & & 16 & 40.46 \\
    & & 32 & 64.95 & & & 32 & 41.22 \\\cline{2-4}\cline{6-8}
    & \multirow{3}{*}{Yes} & 8 & 24.27 & & \multirow{3}{*}{Yes} & 8 & 14.86\\
    & & 16 & 34.98 & & & 16 & 21.82 \\
    & & 32 & 64.54 & & & 32 & 40.93 \\
\end{tabular}
}
\end{table}

\subsection{Speedup breakdown}

\begin{figure}[h!]
    \centering
    \includegraphics[width=0.90\linewidth]{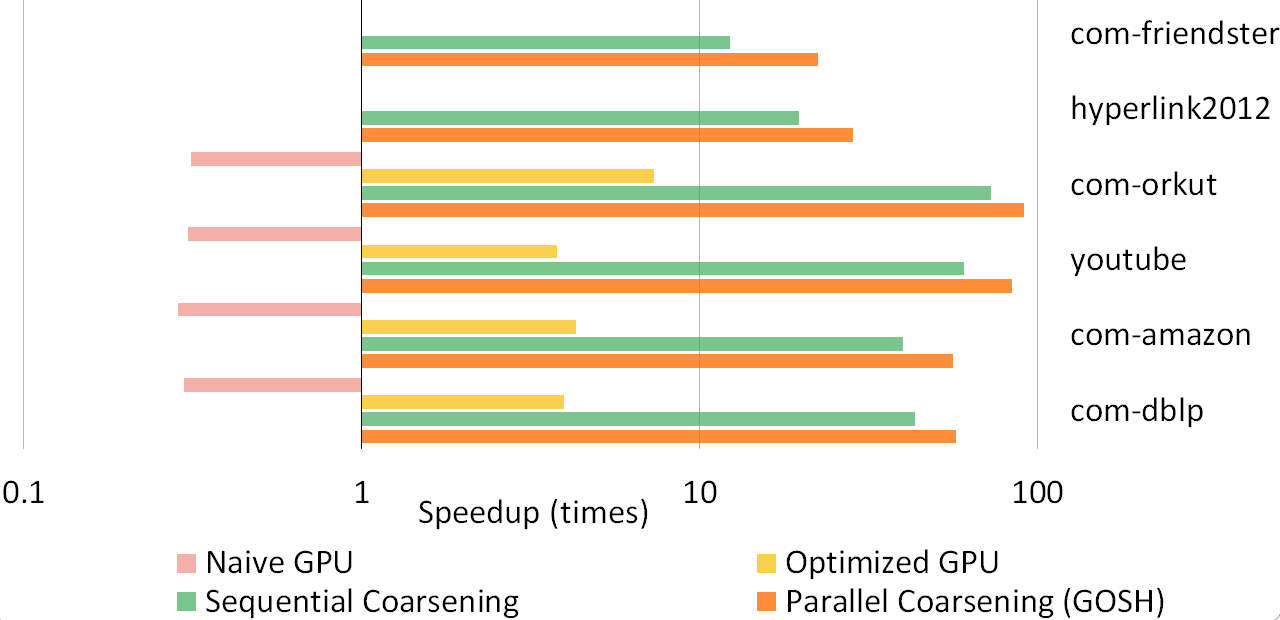}
    \caption{The speedups obtained from running intermediate versions of \malgo compared to our multi-core CPU implementation with 16 threads.}
    \label{fig:speed}
    \Description{The {\em Naive GPU} implementation is slower than a multi-core CPU implementation. The {\em Optimized GPU} implementation is faster than the CPU implementation, the {\em Sequential Coarsening} version is than the optimized version and the {\em Parallel Coarsening} version is the fastest. These observations are consistent for all six graphs.}
\end{figure}

For a more detailed analysis of performance improvements, we run intermediate versions of \malgo and report the speedup over 16-thread CPU implementation. The experiments are conducted with six graphs; two large-scale graphs ({\tt com-friendster}, and {\tt hyperlink2012}), and four medium-scale graphs. We did not run the GPU implementations that are not using coarsening on the large-scale graphs as they take a long amount of time. The results are presented in Figure~\ref{fig:speed}.

The first \malgo version is the {\em Naive GPU} implementation that results in an average slowdown of 3.3$\times$. The {\em Optimized GPU} version leverages architecture-specific optimizations to reduce memory access overhead. That is, global memory is organized to have coalesced accesses, and shared memory is utilized to reduce the number of global memory accesses. This version is 5.4$\times$ faster than the 16-thread one. These two versions do not use coarsening.\looseness=-1 

The next version employs {\em Sequential Coarsening}, as well as all the GPU optimizations from the previous one. This \malgo version scores an average speedup of 45$\times$ over the CPU version, while maintaining the embedding quality as shown in Table~\ref{tab:medium_results}. This is due to the cumulative nature of the updates on the coarsened graphs, where a single update on a super vertex is propagated to all the vertices it contains.\looseness=-1  

The {\em Parallel Coarsening} version, which is the final \malgo, further improves the performance. As discussed in Section~\ref{coarsening-performance}, the performance difference between {\em Parallel Coarsening} and {\em Sequential Coarsening} is expected to be more for larger graphs. For instance, on {\tt com-friendster}, sequential and parallel coarsening phases take 2468.52 and 235.38 seconds, respectively~(Table \ref{table:seq_parallel_coarsening}). On the same graph, the total run-time of \malgo-normal is 2720.82 seconds~(Table \ref{tab:large_results}). In other words, parallel coarsening results in an $80\%$ improvement on performance.

\section{Related Work}\label{sec:rel}
There are various approaches proposed for embedding graphs. A survey on these techniques can be found in~\cite{Goyal2018GraphET}. For instance, in matrix factorization based embedding, the matrix representing the relationship between vertices in the graph is factorized~\cite{roweis2000nonlinear, lem,hope,grarep}. Another approach in the literature is the sampling-based embedding~\cite{deepwalk, node2vec, LINE, verse18} in which the samples are drawn from the graph and used to train a single-layer neural network. Different embedding algorithms use different sampling strategies, most notable of which are random walks~\cite{deepwalk, node2vec}. 
Each algorithm in the literature tries to capture the structural and role/class information based on a similarity measure and/or a sampling strategy. There also have been attempts~\cite{hope, verse18} for a more generalized embedding process. 

Graph embedding is an expensive task and hard to perform without high-performance hardware. 
Graph coarsening \cite{HARP, mile18}, as well as distributed system approaches~\cite{swivel, pbg19, ordentlich2016network} have been previously used to tackle this issue. 
However, these methods do not fully utilize a specialized, now ubiquitous piece of hardware, GPU. The literature does not usually focus on the performance of coarsening since, on CPUs, embedding is slow and coarsening time is negligible. However, this is not the same for GPUs. Furthermore, even state-of-the-art embedding tools are using coarsening schemes that cannot shrink the graphs well.\looseness=-1

There have been attempts to make the embeddings faster by utilizing GPUs; \graphvite generates samples on the host device and performs the embedding on the GPU~\cite{graphvite19}. However, when the total GPU memory is not capable of storing the embedding matrix, the tool cannot perform the embedding. To the best of our knowledge, 
there is no sampling-based graph embedding tool in the literature which is designed for memory restricted but powerful GPUs. \malgo tries to utilize the power of GPUs by applying coarsening and using a judiciously orchestrated CPU-GPU parallelism.\looseness=-1

\section{Conclusion and Future Work} \label{sec:con}

In this paper, we introduce a high-quality, fast graph embedding approach that utilizes CPU and GPU at the same time. The tool can embed any graph by employing a  partitioning schema along with dynamic on-CPU sampling. In addition to this, we provide optimization techniques to minimize GPU idling and maximize GPU utilization during embedding. Furthermore, a parallel coarsening algorithm, which outperforms the state-of-the-art coarsening techniques both in terms of efficiency and speed, is proposed. We fine-tuned the coarsening approach to produce high-quality embeddings at a fraction of the time spent by the state-of-the-art. Our experiments demonstrate the effectiveness of \malgo on a wide variety of graphs.
In the future, we are planning to make \malgo publicly available with easy-to-use interfaces for widely used software such as {\tt Matlab} and {\tt scikit-learn}. Furthermore, we will extend our work for other ML tasks such as classification and anomaly detection.\looseness=-1

\bibliographystyle{ACM-Reference-Format}
\bibliography{main}


\begin{thebibliography}{23}


\ifx \showCODEN    \undefined \def \showCODEN     #1{\unskip}     \fi
\ifx \showDOI      \undefined \def \showDOI       #1{#1}\fi
\ifx \showISBNx    \undefined \def \showISBNx     #1{\unskip}     \fi
\ifx \showISBNxiii \undefined \def \showISBNxiii  #1{\unskip}     \fi
\ifx \showISSN     \undefined \def \showISSN      #1{\unskip}     \fi
\ifx \showLCCN     \undefined \def \showLCCN      #1{\unskip}     \fi
\ifx \shownote     \undefined \def \shownote      #1{#1}          \fi
\ifx \showarticletitle \undefined \def \showarticletitle #1{#1}   \fi
\ifx \showURL      \undefined \def \showURL       {\relax}        \fi
\providecommand\bibfield[2]{#2}
\providecommand\bibinfo[2]{#2}
\providecommand\natexlab[1]{#1}
\providecommand\showeprint[2][]{arXiv:#2}

\bibitem[\protect\citeauthoryear{Belkin and Niyogi}{Belkin and Niyogi}{2001}]%
        {lem}
\bibfield{author}{\bibinfo{person}{M. Belkin} {and} \bibinfo{person}{P.
  Niyogi}.} \bibinfo{year}{2001}\natexlab{}.
\newblock \showarticletitle{Laplacian Eigenmaps and Spectral Techniques for
  Embedding and Clustering}. In \bibinfo{booktitle}{\emph{Proc. 14th Int. Conf.
  on Neural Information Processing Systems: Natural and Synthetic}} (British
  Columbia, Canada) \emph{(\bibinfo{series}{NIPS’01})}.
  \bibinfo{publisher}{MIT Press}, \bibinfo{address}{Cambridge, MA, USA},
  \bibinfo{pages}{585–591}.
\newblock


\bibitem[\protect\citeauthoryear{Cao, Lu, and Xu}{Cao et~al\mbox{.}}{2015}]%
        {grarep}
\bibfield{author}{\bibinfo{person}{S. Cao}, \bibinfo{person}{W. Lu}, {and}
  \bibinfo{person}{Q. Xu}.} \bibinfo{year}{2015}\natexlab{}.
\newblock \showarticletitle{GraRep: Learning Graph Representations with Global
  Structural Information}. In \bibinfo{booktitle}{\emph{Proc. 24th ACM Int.
  Conf. on Info. and Knowledge Management}} (Melbourne, Australia)
  \emph{(\bibinfo{series}{CIKM ’15})}. \bibinfo{publisher}{ACM},
  \bibinfo{address}{NY, USA}, \bibinfo{pages}{891–900}.
\newblock


\bibitem[\protect\citeauthoryear{Chen, Perozzi, Hu, and Skiena}{Chen
  et~al\mbox{.}}{2017}]%
        {HARP}
\bibfield{author}{\bibinfo{person}{H. Chen}, \bibinfo{person}{B. Perozzi},
  \bibinfo{person}{Y. Hu}, {and} \bibinfo{person}{S. Skiena}.}
  \bibinfo{year}{2017}\natexlab{}.
\newblock \bibinfo{title}{HARP: Hierarchical Representation Learning for
  Networks}.
\newblock
\newblock
\showeprint[arxiv]{1706.07845}~[cs.SI]


\bibitem[\protect\citeauthoryear{Fawcett}{Fawcett}{2006}]%
        {roc}
\bibfield{author}{\bibinfo{person}{T. Fawcett}.}
  \bibinfo{year}{2006}\natexlab{}.
\newblock \showarticletitle{An Introduction to ROC Analysis}.
\newblock \bibinfo{journal}{\emph{Pattern Recogn. Lett.}} \bibinfo{volume}{27},
  \bibinfo{number}{8} (\bibinfo{date}{June} \bibinfo{year}{2006}),
  \bibinfo{pages}{861–874}.
\newblock
\showISSN{0167-8655}
\urldef\tempurl%
\url{https://doi.org/10.1016/j.patrec.2005.10.010}
\showDOI{\tempurl}


\bibitem[\protect\citeauthoryear{Goyal and Ferrara}{Goyal and Ferrara}{2018}]%
        {Goyal2018GraphET}
\bibfield{author}{\bibinfo{person}{P. Goyal} {and} \bibinfo{person}{E.
  Ferrara}.} \bibinfo{year}{2018}\natexlab{}.
\newblock \showarticletitle{Graph Embedding Techniques, Applications, and
  Performance: A Survey}.
\newblock \bibinfo{journal}{\emph{Knowl. Based Syst.}}  \bibinfo{volume}{151}
  (\bibinfo{year}{2018}), \bibinfo{pages}{78--94}.
\newblock


\bibitem[\protect\citeauthoryear{Grover and Leskovec}{Grover and
  Leskovec}{2016}]%
        {node2vec}
\bibfield{author}{\bibinfo{person}{A. Grover} {and} \bibinfo{person}{J.
  Leskovec}.} \bibinfo{year}{2016}\natexlab{}.
\newblock \bibinfo{title}{node2vec: Scalable Feature Learning for Networks}.
\newblock
\newblock
\showeprint[arxiv]{1607.00653}~[cs.SI]


\bibitem[\protect\citeauthoryear{{Hu}, {Aggarwal}, {Ma}, and {Huai}}{{Hu}
  et~al\mbox{.}}{2016}]%
        {anomaly_detection}
\bibfield{author}{\bibinfo{person}{R. {Hu}}, \bibinfo{person}{C.~C.
  {Aggarwal}}, \bibinfo{person}{S. {Ma}}, {and} \bibinfo{person}{J. {Huai}}.}
  \bibinfo{year}{2016}\natexlab{}.
\newblock \showarticletitle{An embedding approach to anomaly detection}. In
  \bibinfo{booktitle}{\emph{2016 IEEE 32nd Int. Conf. on Data Eng. (ICDE)}}.
  \bibinfo{pages}{385--396}.
\newblock


\bibitem[\protect\citeauthoryear{Lerer, Wu, Shen, Lacroix, Wehrstedt, Bose, and
  Peysakhovich}{Lerer et~al\mbox{.}}{2019}]%
        {pbg19}
\bibfield{author}{\bibinfo{person}{A. Lerer}, \bibinfo{person}{L. Wu},
  \bibinfo{person}{J. Shen}, \bibinfo{person}{T. Lacroix}, \bibinfo{person}{L.
  Wehrstedt}, \bibinfo{person}{A. Bose}, {and} \bibinfo{person}{A.
  Peysakhovich}.} \bibinfo{year}{2019}\natexlab{}.
\newblock \bibinfo{title}{PyTorch-BigGraph: A Large-scale Graph Embedding
  System}.
\newblock
\newblock
\showeprint[arxiv]{1903.12287}~[cs.LG]


\bibitem[\protect\citeauthoryear{Leskovec and Krevl}{Leskovec and
  Krevl}{2014}]%
        {snapnets}
\bibfield{author}{\bibinfo{person}{Jure Leskovec} {and} \bibinfo{person}{Andrej
  Krevl}.} \bibinfo{year}{2014}\natexlab{}.
\newblock \bibinfo{title}{{SNAP Datasets}: {Stanford} Large Network Dataset
  Collection}.
\newblock \bibinfo{howpublished}{\url{http://snap.stanford.edu/data}}.
\newblock


\bibitem[\protect\citeauthoryear{Liang, Gurukar, and Parthasarathy}{Liang
  et~al\mbox{.}}{2018}]%
        {mile18}
\bibfield{author}{\bibinfo{person}{J. Liang}, \bibinfo{person}{S. Gurukar},
  {and} \bibinfo{person}{S. Parthasarathy}.} \bibinfo{year}{2018}\natexlab{}.
\newblock \bibinfo{title}{MILE: A Multi-Level Framework for Scalable Graph
  Embedding}.
\newblock
\newblock
\showeprint[arxiv]{1802.09612}~[cs.AI]


\bibitem[\protect\citeauthoryear{Liben-Nowell and Kleinberg}{Liben-Nowell and
  Kleinberg}{2003}]%
        {linkprediction}
\bibfield{author}{\bibinfo{person}{D. Liben-Nowell} {and} \bibinfo{person}{J.
  Kleinberg}.} \bibinfo{year}{2003}\natexlab{}.
\newblock \showarticletitle{The Link Prediction Problem for Social Networks}.
  In \bibinfo{booktitle}{\emph{Proc. 12th Int. Conf. on Information and
  Knowledge Management}} (New Orleans, LA, USA) \emph{(\bibinfo{series}{CIKM
  ’03})}. \bibinfo{publisher}{ACM}, \bibinfo{address}{NY, USA},
  \bibinfo{pages}{556–559}.
\newblock


\bibitem[\protect\citeauthoryear{Meusel}{Meusel}{2015}]%
        {hl}
\bibfield{author}{\bibinfo{person}{R. Meusel}.}
  \bibinfo{year}{2015}\natexlab{}.
\newblock \showarticletitle{The Graph Structure in the Web – Analyzed on
  Different Aggregation Levels}.
\newblock \bibinfo{journal}{\emph{Journal of Web Science}}  \bibinfo{volume}{1}
  (\bibinfo{date}{08} \bibinfo{year}{2015}), \bibinfo{pages}{33--47}.
\newblock


\bibitem[\protect\citeauthoryear{Mislove, Marcon, Gummadi, Druschel, and
  Bhattacharjee}{Mislove et~al\mbox{.}}{2007}]%
        {yt}
\bibfield{author}{\bibinfo{person}{A. Mislove}, \bibinfo{person}{M. Marcon},
  \bibinfo{person}{K.~P. Gummadi}, \bibinfo{person}{P. Druschel}, {and}
  \bibinfo{person}{B. Bhattacharjee}.} \bibinfo{year}{2007}\natexlab{}.
\newblock \showarticletitle{Measurement and Analysis of Online Social
  Networks}. In \bibinfo{booktitle}{\emph{Proceedings of the 7th ACM SIGCOMM
  Conference on Internet Measurement}} (San Diego, California, USA)
  \emph{(\bibinfo{series}{IMC ’07})}. \bibinfo{publisher}{ACM},
  \bibinfo{address}{New York, NY, USA}, \bibinfo{pages}{29–42}.
\newblock
\showISBNx{9781595939081}


\bibitem[\protect\citeauthoryear{Niu, Recht, Re, and Wright}{Niu
  et~al\mbox{.}}{2011}]%
        {hogwild}
\bibfield{author}{\bibinfo{person}{F. Niu}, \bibinfo{person}{B. Recht},
  \bibinfo{person}{C. Re}, {and} \bibinfo{person}{Stephen~J. Wright}.}
  \bibinfo{year}{2011}\natexlab{}.
\newblock \showarticletitle{HOGWILD! A Lock-Free Approach to Parallelizing
  Stochastic Gradient Descent}. In \bibinfo{booktitle}{\emph{Proc. 24th Int.
  Conf. on Neural Information Processing Systems}} (Granada, Spain)
  \emph{(\bibinfo{series}{NIPS’11})}. \bibinfo{publisher}{Curran Associates
  Inc.}, \bibinfo{address}{NY, USA}, \bibinfo{pages}{693–701}.
\newblock
\showISBNx{9781618395993}


\bibitem[\protect\citeauthoryear{Ordentlich, Yang, Feng, Cnudde, Grbovic,
  Djuric, Radosavljevic, and Owens}{Ordentlich et~al\mbox{.}}{2016}]%
        {ordentlich2016network}
\bibfield{author}{\bibinfo{person}{E. Ordentlich}, \bibinfo{person}{L. Yang},
  \bibinfo{person}{A. Feng}, \bibinfo{person}{P. Cnudde}, \bibinfo{person}{M.
  Grbovic}, \bibinfo{person}{N. Djuric}, \bibinfo{person}{V. Radosavljevic},
  {and} \bibinfo{person}{Gavin Owens}.} \bibinfo{year}{2016}\natexlab{}.
\newblock \showarticletitle{Network-efficient distributed word2vec training
  system for large vocabularies}. In \bibinfo{booktitle}{\emph{Proc. 25th ACM
  Int. Conf. on Information and Knowledge Management}}.
  \bibinfo{publisher}{ACM}, \bibinfo{address}{xx}, \bibinfo{pages}{1139--1148}.
\newblock


\bibitem[\protect\citeauthoryear{Ou, Cui, Pei, Zhang, and Zhu}{Ou
  et~al\mbox{.}}{2016}]%
        {hope}
\bibfield{author}{\bibinfo{person}{M. Ou}, \bibinfo{person}{P. Cui},
  \bibinfo{person}{J. Pei}, \bibinfo{person}{Z. Zhang}, {and}
  \bibinfo{person}{W. Zhu}.} \bibinfo{year}{2016}\natexlab{}.
\newblock \showarticletitle{Asymmetric Transitivity Preserving Graph
  Embedding}. In \bibinfo{booktitle}{\emph{Proc. 22nd Int. Conf. on Knowledge
  Discovery and Data Mining}} (San Francisco, CA, USA)
  \emph{(\bibinfo{series}{KDD ’16})}. \bibinfo{publisher}{ACM},
  \bibinfo{address}{NY, USA}, \bibinfo{pages}{1105–1114}.
\newblock


\bibitem[\protect\citeauthoryear{Perozzi, Al-Rfou, and Skiena}{Perozzi
  et~al\mbox{.}}{2014}]%
        {deepwalk}
\bibfield{author}{\bibinfo{person}{B. Perozzi}, \bibinfo{person}{R. Al-Rfou},
  {and} \bibinfo{person}{S. Skiena}.} \bibinfo{year}{2014}\natexlab{}.
\newblock \showarticletitle{DeepWalk: Online Learning of Social
  Representations}. In \bibinfo{booktitle}{\emph{Proc. 20th ACM SIGKDD Int.
  Conf. on Knowledge Discovery and Data Mining}} (NY, USA)
  \emph{(\bibinfo{series}{KDD ’14})}. \bibinfo{publisher}{ACM},
  \bibinfo{address}{NY, USA}, \bibinfo{pages}{701–710}.
\newblock


\bibitem[\protect\citeauthoryear{Rossi and Ahmed}{Rossi and Ahmed}{2015}]%
        {nr}
\bibfield{author}{\bibinfo{person}{R.~A. Rossi} {and} \bibinfo{person}{N.~K.
  Ahmed}.} \bibinfo{year}{2015}\natexlab{}.
\newblock \showarticletitle{The Network Data Repository with Interactive Graph
  Analytics and Visualization}. In \bibinfo{booktitle}{\emph{AAAI}}.
\newblock
\urldef\tempurl%
\url{http://networkrepository.com}
\showURL{%
\tempurl}


\bibitem[\protect\citeauthoryear{Roweis and Saul}{Roweis and Saul}{2000}]%
        {roweis2000nonlinear}
\bibfield{author}{\bibinfo{person}{S.~T. Roweis} {and} \bibinfo{person}{L.~K.
  Saul}.} \bibinfo{year}{2000}\natexlab{}.
\newblock \showarticletitle{Nonlinear dimensionality reduction by locally
  linear embedding}.
\newblock \bibinfo{journal}{\emph{science}} \bibinfo{volume}{290},
  \bibinfo{number}{5500} (\bibinfo{year}{2000}), \bibinfo{pages}{2323--2326}.
\newblock


\bibitem[\protect\citeauthoryear{Shazeer, Doherty, Evans, and Waterson}{Shazeer
  et~al\mbox{.}}{2016}]%
        {swivel}
\bibfield{author}{\bibinfo{person}{N. Shazeer}, \bibinfo{person}{R. Doherty},
  \bibinfo{person}{C. Evans}, {and} \bibinfo{person}{Chris Waterson}.}
  \bibinfo{year}{2016}\natexlab{}.
\newblock \bibinfo{title}{Swivel: Improving Embeddings by Noticing What's
  Missing}.
\newblock
\newblock
\showeprint[arxiv]{1602.02215}~[cs.CL]


\bibitem[\protect\citeauthoryear{Tang, Qu, Wang, Zhang, Yan, and Mei}{Tang
  et~al\mbox{.}}{2015}]%
        {LINE}
\bibfield{author}{\bibinfo{person}{J. Tang}, \bibinfo{person}{M. Qu},
  \bibinfo{person}{M. Wang}, \bibinfo{person}{M. Zhang}, \bibinfo{person}{J.
  Yan}, {and} \bibinfo{person}{Q. Mei}.} \bibinfo{year}{2015}\natexlab{}.
\newblock \showarticletitle{LINE: Large-Scale Information Network Embedding}.
  In \bibinfo{booktitle}{\emph{Proc. 24th Int. Conf. on World Wide Web}}
  (Florence, Italy). \bibinfo{publisher}{IW3C2}, \bibinfo{pages}{1067–1077}.
\newblock


\bibitem[\protect\citeauthoryear{Tsitsulin, Mottin, Karras, and
  M\"{u}ller}{Tsitsulin et~al\mbox{.}}{2018}]%
        {verse18}
\bibfield{author}{\bibinfo{person}{A. Tsitsulin}, \bibinfo{person}{D. Mottin},
  \bibinfo{person}{P. Karras}, {and} \bibinfo{person}{E. M\"{u}ller}.}
  \bibinfo{year}{2018}\natexlab{}.
\newblock \showarticletitle{VERSE: Versatile Graph Embeddings from Similarity
  Measures}. In \bibinfo{booktitle}{\emph{Proc. World Wide Web Conference}}
  (Lyon, France) \emph{(\bibinfo{series}{WWW ’18})}.
  \bibinfo{publisher}{IW3C2}, \bibinfo{address}{Republic and Canton of Geneva,
  CHE}, \bibinfo{pages}{539–548}.
\newblock


\bibitem[\protect\citeauthoryear{Zhu, Xu, Tang, and Qu}{Zhu
  et~al\mbox{.}}{2019}]%
        {graphvite19}
\bibfield{author}{\bibinfo{person}{Z. Zhu}, \bibinfo{person}{S. Xu},
  \bibinfo{person}{J. Tang}, {and} \bibinfo{person}{M. Qu}.}
  \bibinfo{year}{2019}\natexlab{}.
\newblock \showarticletitle{GraphVite: A High-Performance CPU-GPU Hybrid System
  for Node Embedding}. In \bibinfo{booktitle}{\emph{The World Wide Web
  Conference}} (CA, USA) \emph{(\bibinfo{series}{WWW ’19})}.
  \bibinfo{publisher}{ACM}, \bibinfo{address}{NY, USA},
  \bibinfo{pages}{2494–2504}.
\newblock


\end{thebibliography}

\end{document}